\documentstyle[epsfig]{article}
\input{epsf}
\begin{document}

\thispagestyle{empty}


\begin{flushright}
  BUTP-99/26\\BUHE-99-08
\end{flushright}

\vskip 25mm

\begin{center}

{\Huge \bf HadAtom99}

\vskip 7mm

Workshop on Hadronic Atoms\footnote{Supported in part by the Swiss National 
Science Foundation, and by TMR, BBW-Contract No. 97.0131 
and EC-Contract No. ERBFMRX-CT980169 (EURODA$\Phi$NE).}\\
Institut f\"ur Theoretische Physik, Universit\"at Bern\\
Sidlerstrasse 5, CH-3012 Bern, Schweiz\\
 October 14 - 15, 1999\\
\end{center}

\vskip 20mm
\begin{center}
{\large J\"urg Gasser} \\
   {\it Institut f\"ur Theoretische  Physik, Universit\"at Bern,  
   CH-3012 Bern, Schweiz}
 \vskip 5mm
{\large Akaki Rusetsky} \\
   {\it Institut f\"ur Theoretische  Physik, Universit\"at Bern,  
   CH-3012 Bern, Schweiz}\\
   {\it Bogoliubov Laboratory of Theoretical Physics, JINR, 141980 Dubna, Russia}\\
   {\it HEPI, Tbilisi State University, 380036 Tbilisi, Georgia}
 \vskip 5mm
and 
\vskip 5mm
{\large J\"urg Schacher} \\
   {\it Laboratorium f\"ur Hochenergiephysik, Universit\"at Bern,
   CH-3012 Bern, Schweiz}
\end{center}

\vskip 20mm
\begin{abstract}
\baselineskip 1.5em
These are the proceedings of the workshop "HadAtom99", held at
the Institut f\"ur Theoretische Physik, Universit\"at Bern,
 October 14 - 15, 1999.
 The main topics  
 discussed at the  workshop were the physics of hadronic atoms and in 
this context recent results from Chiral Perturbation Theory.
Included here are the program, the list of participants and a short
contribution from the speakers.
\end{abstract}

\clearpage

\section{Introduction} 
There has been a growth of interest in hadronic atoms in recent
years. On the experimental side, energy levels of hadronic atoms
as well as  their  decay into neutral hadrons have been or will be
measured with high precision: Pionic hydrogen and $\pi^-d$ atoms at PSI,
 pionium  at CERN (DIRAC), and $K^-p$ as well as $K^-d$ atoms 
at DAFNE (DEAR).
On the theoretical side, effective lagrangian techniques have
shown to be very efficient also here: Bound state calculations --
 in particular the determination of energy levels and decay widths -- can now
be performed with surprising  ease.

For these reasons, Leonid  Nemenov suggested to organize a workshop on
 this subject at the Institut f\"ur Theoretische Physik, Universit\"at
 Bern, similarly to the one organized at  
 Dubna
(Russia) in May 1998 on the same subject. The talks were devoted to recent
experimental and theoretical progress in the investigations of hadronic
 atoms.

The meeting had 48 participants whose names, institutes and e-mail
addresses are listed  below. Results were presented in 25 minutes talks by 26
physicists.
A one page abstract of these presentations as well as  a list of the
most relevant references  has been provided by most of the  speakers. 
These contributions are collected here.
Below we display the list of participants with their e-mail,
then follows the program of the workshop.
Finally, we provide the individual excerpts
submitted by the speakers \footnote{
This kind of miniproceedings had been collected
and posted on the archive in recent years by Hans Bijnens and Ulf 
Mei{\ss}ner at several occasions, see e.g. 
 J. Bijnens and Ulf-G. Mei\ss ner,  
{\rm Chiral Effective Theories}, 205. WE-Heraeus
Seminar, Bad Honnef, November 30 - December 4, 1998, Miniproceedings,
hep-ph/9901381. We found this an
excellent way to summarize the event: 
it does not demand excessive  labor from the authors,  yet
it provides a lot of useful information. Here we follow  
 this idea.}.

 We would like to thank 
all participants for their effort to
 travel to Bern and  
for making this an exciting and lively meeting, and 
our secretaries,   Ruth
 Bestgen and Ottilia H\"anni,  for
 their excellent support.
 Last but not least, we
 thank our colleagues from the organizing committee
 (Heinrich Leutwyler, 
 Valery Lyubovitskij, 
 Leonid Nemenov, 
 Hagop Sazdjian, and  
 Dirk Trautmann)
 for their invaluable help in structuring the meeting.

\vspace*{3cm}

Bern, November 1999

\vspace{1cm}

J\"urg Gasser, Akaki Rusetsky and J\"urg Schacher

\clearpage


\section{List of participants}

\vspace*{.3cm}

{\sf

\begin{center}
\begin{tabular}{l l}
Afanasyev Leonid (CERN-Dubna)          & leonid.afanasev@cern.ch       \\
Amoros Gabriel (Lund)                  & amoros@thep.lu.se             \\
Antonelli Vito (Milan)                 & vito.antonelli@mi.infn.it     \\
Bargholtz Christoph (Stockholm)        & bargholtz@physto.se           \\
Baur Gerhard (FZ Juelich)              & g.baur@fz-juelich.de          \\
Becher Thomas (Bern)                   & becher@itp.unibe.ch           \\
Bellucci Stefano (LNF-INFN)            & bellucci@lnf.infn.it          \\
Colangelo Gilberto (Z\"{u}rich)        & gilberto@physik.unizh.ch      \\
Cugnon Joseph (Liege)                  & j.cugnon@ulg.ac.be            \\
Descotes Sebastien (Orsay)             & descotes@ipno.in2p3.fr        \\
de Simone Patrizia (LNF-INFN)          & patrizia.desimone@lnf.infn.it \\
Drijard Daniel (CERN)                  & daniel.drijard@cern.ch        \\
Eiras Dolors (Barcelona)               & dolors@ecm.ub.es              \\
Gallas Manuel (CERN)                   & gallas@gaes.usc.es            \\
Gashi Agim (Z\"{u}rich)                & gashi@physik.unizh.ch         \\
Gasser Juerg (Bern)                    & gasser@itp.unibe.ch           \\
Girlanda Luca (Orsay)                  & girlanda@ipno.in2p3.fr        \\
Guaraldo Carlo (Frascati)              & carlo.guaraldo@cern.ch        \\
Hasenfratz Peter (Bern)                & hasenfra@itp.unibe.ch         \\
Heim Thomas (Basel)                    & thomas.heim@unibas.ch         \\
Hencken Kai (Basel)                    & k.hencken@unibas.ch           \\
Jensen Thomas (PSI)                    & thomas.jensen@psi.ch          \\
Kaiser Roland (Bern)                   & kaiser@itp.unibe.ch           \\
Knecht Marc (Marseille)                & knecht@cptsu5.univ-mrs.fr     \\
Lauss Bernhard (LNF-INFN)              & bernhard.lauss@lnf.infn.it    \\
Lednicky Richard (Prague)              & lednicky@fzu.cz               \\
Leutwyler Heinrich (Bern)              & leutwyle@itp.unibe.ch         \\
Lyubovitskij Valery (Dubna-Tomsk)      & lubovit@itp.unibe.ch          \\
Markushin Valery (PSI)                 & markushin@psi.ch              \\
Mei{\ss}ner Ulf (FZ Juelich)           & ulf-g.meissner@fz-juelich.de  \\
Nemenov Leonid (CERN-Dubna)            & leonid.nemenov@cern.ch        \\
Niedermayer Ferenc (Bern)              & niederma@itp.unibe.ch         \\
Pentia Mircea (CERN-Bucharest)         & pentia@ifin.nipne.ro          \\
Petrascu Catalina (LNF-INFN)           & catalina.oana.petrascu@cern.ch\\
Pislak Stefan (BNL)                    & pislak@bnldag.ags.bnl.gov     \\
Ravndal Finn (Oslo)                    & finn.ravndal@fys.uio.no       \\
Rodriguez Ana (CERN)                   & ana.maria.rodriguez.fernandez@cern.ch\\
Rusetsky Akaki (Bern-Dubna-Tbilisi)    & rusetsky@itp.unibe.ch         \\
Santamarina Cibran (CERN)              & cibran@sungaes1.gaes.usc.es   \\
Sazdjian Hagop (Orsay)                 & sazdjian@ipno.in2p3.fr        \\
Schacher Juerg (Bern)                  & schacher@lhep.unibe.ch        \\
Schweizer Julia (Bern)                 & schweize@itp.unibe.ch         \\
Simons Leopold (PSI)                   & leopold.simons@psi.ch         \\
Soto Joan (Barcelona)                  & soto@ecm.ub.es                \\
Stern Jan (Orsay)                      & stern@ipno.in2p3.fr           \\
Tarasov Alexander (Regensburg-Dubna)   & avt@dxndh1.mpi-hd.mpg.de      \\
Trautmann Dirk (Basel)                 & trautmann@ubaclu.unibas.ch    \\
Voskressenskaia Olga (Heidelberg-Dubna)& olga@dxnhd5.mpi-hd.mpg.de     \\
\end{tabular}
\end{center}

}


\newpage 

\section{HadAtom99: Program}

\vspace*{1.cm}

\centerline{\sc Thursday, October 14}

\vspace*{1.cm}

{\sf

\noindent
\begin{tabular}{r c r l l}
10:00 & - & 10:05 & J.~Gasser (Bern)     & Welcome\\ 
&&&&\\
10:05 & - & 10:30 & J.~Schacher (Bern)   & The DIRAC experiment at CERN \\
&&&&\\
10:30 & - & 10:55 & R.~Lednicky (Prague) & Finite size effects on $\pi^+\pi^-$ production in\\ 
      &   &       &                      & continuous and discrete spectrum \\
&&&&\\
10:55 & - & 11:20 & A.V.~Tarasov (Dubna) & Analytical approach to calculation \\
      &   &       &                      & of the $\pi^+\pi^-$ atom production \\
      &   &       &                      & rate in different quantum states \\
&&&&\\
11:20 & - & 11:45 & Coffee&\\ 
&&&&\\
11:45 & - & 12:10 & F.~Ravndal (Oslo)   & Corrections to the pionium lifetime \\ 
&&&&\\
12:10 & - & 12:35 & J.~Soto (Barcelona) & Effective field theory approach to pionium \\ 
12:35 & - & 13:00 & H.~Sazdjian (Orsay) & Pionium lifetime in Generalized Chiral \\
      &   &       &                     & Perturbation Theory \\
&&&&\\
13:00 & - & 14:30 & Lunch&\\ 
&&&&\\
14:30 & - & 14:55 & V.E.~Lyubovitskij (Dubna) & $\pi^+\pi^-$ atom in QCD\\ 
&&&&\\
14:55 & - & 15:20 & A.~Rusetsky (Bern)        & Spectrum and decays of hadronic atoms\\ 
15:20 & - & 15:45 & L.G.~Afanasyev (Dubna)    & Calculation of the breakup probability of \\
      &   &       &                           & $\pi^+\pi^-$ atom in a target with a high accuracy\\ 
&&&&\\
15:45 & - & 16:15 & Coffee&\\ 
&&&&\\
16:15 & - & 17:15 & 
$
\left.
\begin{array}{l}
{\mbox{D.~Trautmann (Basel)}}\\
{\mbox{Th.~Heim (Basel)}}\\
{\mbox{K.~Hencken (Basel)}}\\
\end{array}
\right\}
$
&
$
\left\{
\begin{array}{l}
{\mbox{Group~report}}\\
{\mbox{about~pionium}}\\
{\mbox{interaction~with~matter}}\\
\end{array}
\right.
$ 
\\ 
&&&&\\
17:15 & - & 17:35 & Coffee&\\ 
&&&&\\
17:35 & - & 18:00 & G.~Baur (FZ J\"{u}lich) &  Another exotic relativistic atom: antihydrogen \\
&&&&\\
18:00 & - & 18:25 & J.~Cugnon (Liege) & Multiphoton exchange effects in \\
      &   &       &                   & pionium-matter interaction \\ 
\end{tabular}

}

\newpage

\centerline{\sc Friday, October 15}

\vspace*{1.cm}

{\sf

\noindent
\begin{tabular}{r c r l l}
 9:00 & - &  9:25 & L.L.~Nemenov (CERN-Dubna) & Development in the theoretical dimesoatom \\ 
      &   &       &                           & description required by DIRAC experiment \\
&&&&\\
 9:25 & - &  9:50 & C.~Guaraldo (LNF-INFN)    & The DEAR experiment at DA$\Phi$NE \\ 
&&&&\\
 9:50 & - & 10:15 & L.~Simons (PSI)           & Pionic hydrogen: status and outlook \\ 
&&&&\\
10:15 & - & 10:40 & Coffee&\\ 
&&&&\\
10:40 & - & 11:05 & U.-G.~Mei{\ss}ner (FZ J\"{u}lich) & Pion-kaon scattering \\ 
&&&&\\
11:05 & - & 11:30 & H.~Leutwyler (Bern) & Discussion: pion-kaon scattering \\ 
&&&&\\
11:30 & - & 11:45 & Coffee&\\ 
&&&&\\
11:45 & - & 12:10 & P.~de Simone (LNF-INFN) & $K_{l4}$ decays at DA$\Phi$NE\\ 
&&&&\\
12:10 & - & 12:35 & S.~Pislak (BNL)         & A new measurement of the \\
      &   &       &                         &$K^+\rightarrow \pi^+\pi^- e^+ \nu$ decay \\
&&&&\\
12:35 & - & 13:00 & G.~Amoros (Lund)        & $K_{\ell 4}$ decays: a theoretical discussion \\
&&&&\\
13:00 & - & 14:30 & Lunch &\\ 
&&&&\\
14:30 & - & 14:55 & J.~Stern (Orsay) & Chiral phase transitions at zero temperature\\
&&&&\\
14:55 & - & 15:20 & L.~Girlanda (Orsay) & The two-flavor chiral condensate from \\ 
      &   &       &                     & low-energy $\pi\pi$ phase-shifts \\
&&&&\\
15:20 & - & 15:45 & M.~Knecht (Marseille) & Radiative corrections in $K_{l4}$ decays \\ 
&&&&\\
15:45 & - & 16:15 & Coffee &\\ 
&&&&\\
16:15 & - & 16:40 & G. Colangelo (Zurich) & Numerical solutions of Roy equations\\ 
&&&&\\
16:40 &  &  & J. Gasser (Bern) & Closing remarks \\ 
\end{tabular}

}

\newpage 

\setcounter{equation}{0}

\begin{center}
{\Large\bf The DIRAC Experiment at CERN}

\bigskip

DIRAC Collaboration\\
J. Schacher\\

{\it Laboratorium f\"ur Hochenergiephysik, Universit\"at Bern,} 
{\it CH-3012 Bern, Switzerland}

\end{center}
\bigskip

The experiment DIRAC [1], a magnetic double arm spectrometer, 
aims to measure the $\pi^+\pi^-$ atom lifetime with $10\%$ 
precision, using the high intensity 24~GeV/c 
proton beam of the CERN Proton Synchrotron. Since the value 
of this lifetime of order $10^{-15}$~s is dictated by 
strong interaction at low energy, a precise measurement of 
this quantity enables to study characteristic pion parameters 
in detail and to submit predictions of QCD to a severe check.

After tuning the primary proton beam as well as all detector 
components in detail, the experiment started to 
accumulate data this summer. Preliminary results concerning 
quality and reliability of the setup are presented in 
the following two figures:
  
\bigskip
\begin{figure}[h] 
\begin{center}

\begin{minipage}[t]{7cm}
\mbox{\epsfig{figure=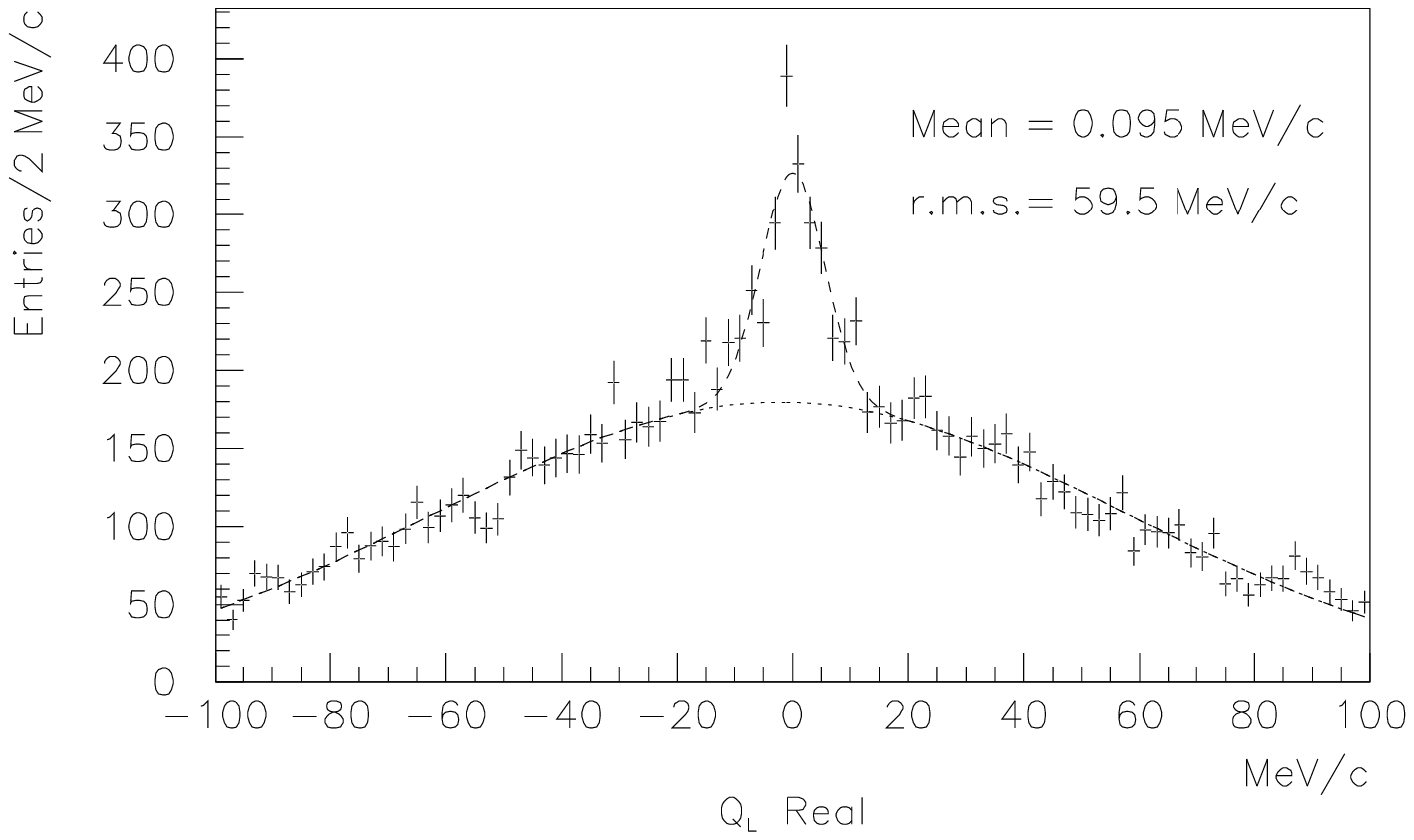,width=7cm}}
\caption{Correlated pairs of $\pi^+$ and 
$\pi^-$ with momenta $p_{lab}$ smaller than 
4.5~GeV/c and $Q_{T}<~4$~MeV/c.}
\end{minipage}
\hspace{2cm}
\begin{minipage}[t]{7cm}
\mbox{\epsfig{figure=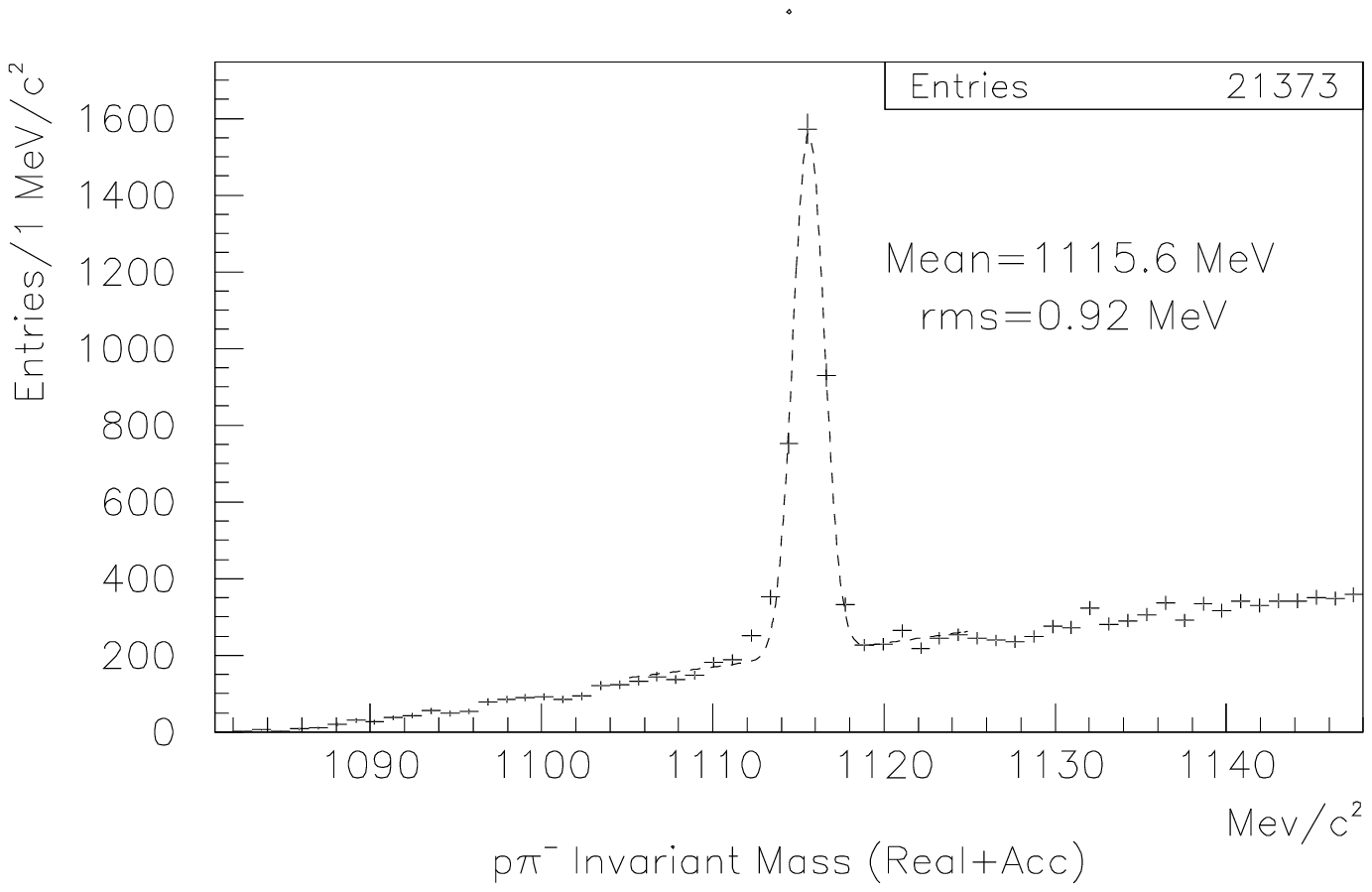,width=7cm}}
\caption{Invariant pair mass of $\pi^-$ and 
p with momentum $p_{lab}$ larger than 3~GeV/c.}
\end{minipage}

\end{center}
\end{figure}
\bigskip

In Fig.~1, the distribution of $\pi^+\pi^-$ pairs in 
dependence of $Q_{L}$ shows a noticeable peak in the yield of 
$\pi^+\pi^-$ pairs with momenta $Q$ less than 5 MeV/c, due to 
Coulomb attraction in the final state. The longitudinal 
component $Q_{L}$ is defined as the projection 
of the relative momentum $\vec{Q}$ in the $\pi^+\pi^-$ 
system on the direction of the $\pi^+\pi^-$ lab momentum. The 
mean value of the peak, $\langle Q_{L} \rangle$~=~0.095~MeV/c, 
is small and hence clear evidence for good setup alignment.

As emphasized in Ref.~1, the resolution in $Q_{L}$ and 
$Q_{T}$ plays a key role in the data analysis. In order 
to keep track of the relative momentum resolution, the 
invariant mass distribution of p and $\pi^-$ 
--- products of the $\Lambda$ decay --- was carefully 
investigated. The position of the mass peak in Fig.~2 
corresponds to $\Lambda$ particles, recorded and 
reconstructed in the magnetic spectrometer. 
The width of the $\Lambda$ peak, mainly given by 
the momentum resolution, is found to be 
less than $1~MeV/{c^{2}}$ (see Fig.~2).
Thus, the DIRAC collaboration is confident, that the 
magnetic field and the location of the track detectors 
are known with adequate precision.

\bigskip\bigskip\bigskip

\noindent [1] B. Adeva {\it et al.}, Proposal to the SPSLC, 
              CERN/SPSLC 95-1 (1995).

\newpage 

\setcounter{equation}{0}

\begin{center} 
{\Large\bf Finite--size effects on $\pi^+\pi^-$ 
production in continuous and discrete spectrum} 
 
\bigskip
 
R.~Lednick\'{y}   
\\ {Institute of Physics, 
Na Slovance 2, 18221 Prague 8, Czech Republic} 
\end{center} 
 
\bigskip 
 
The determination, on a per cent level accuracy, 
of the breakup probability $P_{br}=N_A^{br}/N_A$ of the 
$\pi^+\pi^-$ atoms produced by a high energy beam in a target 
is of principle importance for a precise lifetime measurement of these 
atoms in the experiment DIRAC [1] currently running at CERN.
Clearly, the breakup probability is a unique function of the 
target geometry and material, the Lorentz factor and the ground 
state lifetime.

While the number of the breakup atoms $N_A^{br}$ is measured directly
as an excess of the $\pi^+\pi^-$ pairs with very low relative momentum,
the number $N_A$ of the produced $\pi^+\pi^-$ atoms is calculated
based on the Migdal-Watson theory of final state interaction 
in discrete [2,3] and continuous [4] spectrum. This calculation is
sensitive to the space--time extent of the pion production region,
characterized  by the distance $r^{*}$ between the $\pi^+$ and 
$\pi^-$ production points in their c.m.s..
In ref. [2], the $r^*$--dependence was treated in an approximate way,
dividing the pion sources into short--lived and long--lived ones.
It was assumed that $r^*=0$ for pion pairs arising solely from the
short--lived sources and characterized by the distances $r^*$
much smaller than the Bohr radius of the $\pi^+\pi^-$ system: 
$|a|=387.5$ fm, otherwise $r^*=\infty$.
 
The finite size correction to such calculated number of 
free $\pi^+\pi^-$ pairs 
arises mainly from the region of small relative momenta 
compared with $1/r^*$. It
is determined by the three dimensionless 
combinations $r^*/a$, $f_0/r^*$ and $f_0/a$ 
of $r^*$, $a$  and the s--wave $\pi^+\pi^-$-scattering length: 
$f_0=\frac 13 (2a_0+a_2)\approx 0.23$ fm.
Typically $\langle r^*\rangle \sim$ 10 fm so that the correction
is dominated by the strong interaction effect and can amount up to
$\sim 10\%$.
 
Due to a small binding energy $\epsilon\sim (2\mu a^2)^{-1}$, 
the finite-size correction to the discrete spectrum at 
$r^*\ll |a|$
is nearly the same as that to the continuous spectrum at zero energy. 
Since $N_A$ is actually determined by a ratio of the pions produced 
in discrete
and continuous spectrum, this correction, up to a fraction of $r^*/a$, 
would cancel out in the breakup probability 
provided we could measure the number of free $\pi^+\pi^-$ pairs 
in the region of very small relative momenta, much less than 
$1/r^*$ [5,6]. 
With the increasing 
relative momentum the correction to the production of the free pairs 
decreases, the decrease 
being  faster for  pions emitted at 
larger distances $r^{*}$. Particularly important is the case when one of the 
two pions comes from the $\omega$-meson decay and $r^{*}\sim 30$ fm. 
As a result, in DIRAC conditions, we expect 
the net finite size correction in the breakup probability 
of the order of per cent. It leads to several per cent effect in the
measured lifetime and, in first approximation, can be neglected
compared with the expected statistical error of $\sim 10\%$.
 
We discuss how 
to diminish the systematic error 
related to the finite size effects, using 
the correlation data on identical charged pions (containing the 
information about the distances $r^*$ between the 
pion production points in the same experiment) 
together with the complete phase--space 
simulations within transport models.
We also show that the usual equal--time approximation, the neglect 
of the space--time coherence as well as of the
transition $\pi^+\pi^- \leftrightarrow \pi^0\pi^0$ and of the residual
nucleus charge introduce negligible systematic errors of the order
of per mill in the measured lifetime.

\bigskip 
 
\noindent [1] B.~Adeva et al.: Lifetime measurement of 
$\pi^+\pi^-$ atoms to test low energy QCD predictions,\\ 
Proposal to the SPSLC, CERN/SPSLC 95-1, updated 10 November 1995. 
 
\noindent [2] L.~Nemenov: 
{\it Yad.~Fiz.} {\bf 41} (1985) 980. 
 
\noindent [3] V.L.~Lyuboshitz: 
{\it Yad.~Fiz.} {\bf 48} (1988) 1501 ({\it 
Sov.~J.~Nucl.~Phys.} {\bf 48} (1988) 956).

\noindent [4]  R.~Lednicky, V.L.~Lyuboshitz: 
{\it Yad.~Fiz.} {\bf 35} (1982) 1316 ({\it Sov.~J.~Nucl.~Phys.} 
{\bf 35} (1982) 770). 

\noindent [5]  R.~Lednicky: On the breakup probability
of $\pi^+\pi^-$ atoms, DIRAC meeting, February, 1998;\\
Finite-size effects on $\pi^+\pi^-$ production in continuous
and discrete spectrum, DIRAC meeting, February, 1999.

\noindent [6]  L.~Afanasyev, O.~Voskresenskaya: 
{\it Phys.~Lett. B} {\bf 453} (1999) 302.

\newpage 

\setcounter{equation}{0}

   \begin{center}
   {\Large\bf Analytical approach to calculation of the $\pi^+\pi^-$ atom 
   production rate in different quantum states}
  
   \bigskip
   
   {\underline{A.Tarasov}}$^{1,3}$ and O.Voskresenskaya$^{2,3}$
   
   $^1$Institut f\"ur Theoretische Physik,    
   Universit\"at Regensburg, D--93040 Regensburg, Germany\\
   $^2$Max-Plank Institut f\"ur Kernphysik,   
   Postfach 103980, D--69029 Heidelberg, Germany\\
   $^3$Joint Institute for Nuclear Research, 
   141980 Dubna, Moscow Region, Russia
   \end{center}
   
   \bigskip
   
   A method of measurement of the $\pi^+\pi^-$-atom lifetime proposed by
   L.L.Nemenov [1] and which is under realization in the experiment
   DIRAC [2] at CERN Proton Synchrotron is essentially based on the
   assumption that the ratio $R(nS/pS)$ of the
   $\pi^+\pi^-$-atom production rate in $nS$-states in hadron-nuclear
   interactions at high energy to the production rate of so the called
   ``Coulomb'' $\pi^+\pi^-$ pairs (i.e. free $\pi^+\pi^-$ pairs produced at a
   distance less or of the order of the $\pi^+\pi^-$-atom Bohr radius and having
   the relative momentum of an order of the Bohr momentum) can be calculated
   with accuracy better than 1\%. Here we consider the accuracy of this ratio.
   
   According to [1,3], the ratio can be written as:
   \begin{equation}
   \label{e1}
   R(n\mathrm{s}/p\mathrm{s})=\frac{\displaystyle
   \left|\int M(\vec r) \psi_{n\mathrm{s}}(\vec r)\,d^3r \right|^2 }
   {\displaystyle   
   \left|\int M(\vec r) \psi_{p\mathrm{s}}(\vec r)\,d^3r \right|^2 } \, ,
   \end{equation}
   where $\psi_{n\mathrm{s}}(\vec r)$ is the $\pi^+\pi^-$-atom wave function with the
   principal quantum number $n$ and zero orbital one ($n$s-states);
   $\psi_{p\mathrm{s}}(\vec r)$ is the wave function of $\pi^+\pi^-$ pair with
   the relative momentum $p$ and with zero orbital momentum (s-wave);
   $M(\vec{r})$ is the amplitude of $\pi^+\pi^-$-system production at the
   relative distance $\vec r$.
   
   In ref.[4] it is shown that in the first order of the perturbation
   theory over the strong interaction $R(nS/pS)$ is expressed via the a
   squared ratio of the well-known Coulomb wave functions.
   
   Numerical calculations in the resent paper [5] show that the same
   ratio between discrete states is also expressed via the a squared
   ratio of the Coulomb wave functions at origin.
   
   In this paper we analyse the influence of the strong interaction on the
   ratio $R(nS/pS)$ analytically using a square well approximation for the
   strong interaction potential. It is shown that the result of ref.[4]
   is valid with accuracy of an order of ratio of the
   $\pi^+\pi^-$-scattering length to the $\pi^+\pi^-$-atom Bohr radius
   which is approximately equal to $10^{-3}$. Thus, both perturbative and
   nonperturbative calculations of $R(nS/pS)$ confirm that Nemenov's result
   [1] for this quantity is valid with the required accuracy.
   
   \bigskip
   
   \noindent [1] L.L. Nemenov,  {\em Yad.Fiz.} {\bf 41} (1985) 980.
   
   \noindent [2] B. Adeva et al., {\it Lifetime measurement of $\pi^+\pi^-$
   atoms to test low energy QCD predictions}, (Proposal to the SPSLC, CERN/SPSLC
   95--1, SPSLC/P 284, Geneva 1995).
   
   \noindent [3] L.G. Afanasyev, O.O. Voskresenskaya and V.V. Yazkov, {\em JINR
   Communication}, P1--97--306, Dunba 1997.
   
   \noindent [4] L.G. Afanasyev and O.O. Voskresenskaya, {\it Physics Letters B} 
   {\bf 453} (1999) 302.
   \noindent [5] I. Amirkhanov, I. Puzynin, A. Tarasov, O. Voskresenskaya
    and O. Zeinalova {\it Physics Letters B} {\bf 452}  (1999) 1.

\newpage 

\setcounter{equation}{0}

\begin{center}
{\Large\bf Corrections to the pionium lifetime}

\bigskip

Finn Ravndal

Institute of Physics, University of Oslo, N-0316 Oslo, Norway.
\end{center}

\bigskip

The charged pions with mass $m_+$ in the 1S ground state of pionium are bound 
in the Coulomb wavefunction $\Psi(r)$ with characteristic momentum 
$\gamma = \alpha m_+/2$. Its dominant decay mode is into two neutral pions 
with mass $m_0$. Since the mass difference $\Delta m = m_+ - m_0$ is small, 
both the pions in the initial state of the hadronic atom and in the 
final state of the decay are non-relativistic. This was first emphasized by 
Labelle and Buckley [1] who showed that the leading term of the decay rate 
$\Gamma$ can be obtained by calculating the energy shift $\Delta E$ within 
non-relativistic effective field theory using the relation 
$\Gamma = -2\,\mbox{Im}\,\Delta E$. The corresponding Lagrangian was then 
extended by Kong and Ravndal [2] to include higher order interactions which 
allow corrections to the dominant decay rate to be systematically calculated.

In order to determine the unknown coupling constants of the effective 
Lagrangian, one must match the non-relativistic scattering amplitude for 
$\pi^+ + \pi^- \rightarrow \pi^0 + \pi^0$ from this theory to that which 
follows from relativistic, chiral perturbation theory at the same energy. 
At low momenta this latter one is defined to be
\begin{equation}
         T_R(p) = 32\pi(a + b p^2/m_+^2) 
\end{equation}
in the center-of-mass frame where the charged pions have the momentum $p$. The 
scattering length $a$ and the slope parameter $b$ include both higher order 
chiral corrections and isospin-violating effects from the quark mass 
difference $m_u - m_d$ and short-range electromagnetic effects. In the 
matching one must also include the $p^2$ effects coming from the difference 
in normalization of relativistic and non-relativistic states as done correctly
by Gall, Gasser, Lyubovitskij and Rusetsky [3].

At threshold the momentum $p=0$ and the full scattering amplitude with 
long-range Coulomb interactions removed is then given by just this scattering 
length. It can be written in terms of the more conventional isospin-symmetric 
scattering lengths $a_0$ and $a_2$ in the isospin $I=0$ and isospin $I=2$ 
channels as $a = a_0 - a_2 + \Delta a$ where $\Delta a$ includes these 
symmetry-breaking effects. 

Calculating now the transition rate and multiplying by the probability
$|\Psi(0)|^2 = \gamma^3/\pi$ that the pions are found at the same point, one 
obtains the leading order result for the decay rate
\begin{equation}
     \Gamma_0 = {2\over 9}\alpha^3{m_0\over m_+}\sqrt{2\Delta m m_0}\,a^2 
\end{equation}
Relativistic effects in the final state can be obtained from a corresponding 
term of the effective Lagrangian [4] and follows also directly from the 
phase space factor
\begin{equation}
     \Gamma \propto \int\!{d^3k\over (2\pi)^3}\;\delta(2m_+ - 2E_0)
\end{equation}
where $E_0 = m_0 + {k^2/ 2m_0} - {k^4/ 8m_0^3}$ is the energy of the neutral 
pions when we ignore the small binding energy. To this order in 
$\Delta m/m$ the decay rate is thus found to be
\begin{equation}
\Gamma = {2\over 9}\alpha^3\sqrt{2\Delta m m_0}
\left(1+{\Delta m\over 4m}\right) a^2
\end{equation}
This is consistent with the more general result of Gall {\it et al.} [3] and 
has also been found by Eiras and Soto [5].

Remaining corrections can now be obtained by including long-range Coulomb 
interactions and rescattering effects. These have been worked out by  
Gall {\it et al.} [3] and will also follow from the effective theory of 
Eiras and Soto [5]. The above  non-relativistic approach can be similarly 
applied to other hadronic atoms.

\bigskip

\noindent [1] P. Labelle and K. Buckley, hep-ph/9804201.

\noindent [2] X. Kong and F. Ravndal, {\it Phys. Rev.} {\bf D59}, 
014031 (1999). 

\noindent [3] A. Gall, J. Gasser, V.E. Lyubovitskij and A. Rusetsky, 
Phys. Lett. {\bf B462}, 335 (1999).

\noindent [4] X. Kong and F. Ravndal, hep-ph/9905539.

\noindent [5] D. Eiras and J. Soto, hep-ph/9905543.

\newpage 

\setcounter{equation}{0}

\begin{center}
{\Large\bf Effective field theory approach to pionium}

\bigskip

Joan Soto

Departament d'Estructura i Constituents de la Mat\`eria, Universitat de 
Barcelona,
Diagonal 647, 08028 Barcelona, Catalonia, Spain
\end{center}

\bigskip

I shall briefly report on 
 the model-independent approach to pionium 
which has been presented in [1].

\medskip

 It basically consist of identifying the 
relevant energy and momentum scales of the $\pi^{+}\pi^{-}$ atom
 and sequentially integrate them out until we reach 
the lowest relevant scale, namely the binding energy. This is carried out by 
 introducing a series of non-relativistic effective field theories
 and by requiring them to be equivalent at the desired order of accuracy.
Since pionium is an electromagnetic bound state, it contains at least three 
dynamical scales: the mass of the charged pions $m\sim 140 MeV$, the typical 
relative momentum 
in the bound state $m\alpha / 2\sim 0.5 MeV$, and the binding 
energy $m\alpha^2/4\sim 2 keV$ 
[2]. In addition pionium decays mainly to two neutral pions through 
the strong interactions, which brings in an additional energy scale $\Delta m
\sim 5 MeV$, the 
difference between the charge and neutral pion masses, and its associated 
three momentum
$s=\sqrt{2m\Delta m}\sim 40 MeV$. 
 In view of these numerical values two important observations are in order:
(i) all the scales are considerably smaller than $m\sim 140$ MeV, which 
implies that a 
non-relativistic approach should be appropriated, and
(ii) all the scales, including $m$, are considerably smaller than the typical 
hadronic scale, say the rho mass, and hence all the necessary information 
 is contained in the Chiral Lagrangian ($\chi$L)
coupled to electromagnetism. Since at these energies not only
 electromagnetism but also the strong interactions are amenable to 
a perturbative treatment 
(in $\alpha$ and in Chiral Perturbation Theory ($\chi$PT) respectively),
pionium can be addressed in a model independent way.

\medskip 

Let us then start with the $SU(2)\times SU(2)$ $\chi$L coupled to 
electromagnetism. After integrating out the scale $m$ for pion pairs near 
threshold one should obtain what may be called a Non-Relativistic 
$\chi$L (NR$\chi$L). The degrees of freedom of NR$\chi$L are photons and
non-relativistic charged and neutral pion fields. 
NR$\chi$L is analogous to Non-Relativistic QED [2], but contains
non-relativistic pseudoscalar fields instead of Pauli spinor fields and 
includes the effect of the strong interactions. 
We present the general form of the NR$\chi$L including isospin breaking terms.
Since $\Delta m \gg m\alpha/2$ , $m\alpha^2/4$
 we can also integrate out this scale  and its associated 
three momentum $s$ (neutral pions). We obtain
 a second Non-Relativistic $\chi$L which may be abbreviated as 
NR$\chi$L$^{\prime}$.
 Its degrees of freedom are photons and non-relativistic charged pions. 
The introduction of NR$\chi$L$^{\prime}$ avoids having to solve eventually a 
coupled channel problem. Finally, integrating out photons 
of energy or momentum of the order $m\alpha /2$ we obtain what may be called
potential NR$\chi$L$^{\prime}$ (pNR$\chi$L$^{\prime}$), in analogy of
 potential NRQED [3]. At the order we are interested in, pNR$\chi$L$^{\prime}$
contains non-relativistic charged pions 
interacting through a potential (electromagnetic and strong), and it is 
totally equivalent to a suitable quantum mechanical Hamiltonian. 
 
\medskip

A second order quantum mechanical perturbation theory calculation in 
pNR$\chi$L$^{\prime}$ allows us to obtain the decay width and the spectrum
at next to leading order in $\Delta m / m$, $m\alpha^2 / 4\Delta m$ and 
$\alpha$. In order to match this precision a yet-to-be-done two loop 
calculation in the $\chi$L including photons is necessary.  

\medskip

The techniques described above may also be useful for other hadronic atoms. 

\medskip

Related work using non-relativistic effective field theories can be found in
[4].

\bigskip

\noindent [1] D. Eiras and J. Soto, hep-ph/9905543.

\noindent [2]  W.E. Caswell and G.P. Lepage, {\it Phys. Lett.} {\bf
B167} (1986) 437.

\noindent [3] A. Pineda and J. Soto, {\it Phys.Rev.} {\bf D59} (1999) 016005,
hep-ph/9805424.

\noindent [4] P. Labelle and K. Buckley, hep-ph/9804201. 

\noindent $\quad$ X. Kong and F. Ravndal, hep-ph/9805357.

\noindent $\quad$ B. R. Holstein, nucl-th/9901041.

\noindent $\quad$ A. Gall, J. Gasser, V. E. Lyubovitskij and A. Rusetsky,
{\it Phys. Lett.} {\bf B462} (1999) 335.

\noindent $\quad$ X. Kong and F. Ravndal, hep-ph/9905539.

\noindent $\quad$  J. Gasser, V. E. Lyubovitskij and
 A. Rusetsky, hep-ph/9910438.

\newpage 

\setcounter{equation}{0}

\begin{center}
{\Large\bf Pionium lifetime in generalized chiral perturbation theory}

\bigskip

H. Sazdjian

Groupe de Physique Th\'eorique, Institut de Physique Nucl\'eaire,\\
Universit\'e Paris XI, 91406 Orsay Cedex, France 
\end{center}

\bigskip

The relationship between the pionium lifetime and the $\pi\pi$ scattering
lengths, including the sizable electromagnetic corrections, is analyzed
in the framework of generalized chiral perturbation theory, in which
the quark condensate value is left as a free parameter. The variation curve
of the lifetime as a function of the combination $(a_0^0-a_0^2)$ of the
$S$-wave scattering lengths is presented.

\newpage 

\setcounter{equation}{0}

\begin{center}
{\Large\bf $\pi^+\pi^-$ Atom in QCD}

\bigskip

J. Gasser$^1$, \underline{V.E. Lyubovitskij}$^{2,3}$, 
 and A.G. Rusetsky$^{1,2,4}$  

{$^1$ Institute for Theoretical Physics, University of Bern,
Sidlerstrasse 5, CH-3012, Bern, Switzerland}

{$^2$ Bogoliubov Laboratory of Theoretical Physics, Joint Institute
for Nuclear Research, 141980 Dubna, Russia}

{$^3$ Department of Physics, Tomsk State University, 634050 Tomsk, Russia}

{$^4$ HEPI, Tbilisi State University, 380086 Tbilisi, Georgia}
\end{center}

\bigskip

The DIRAC experiment at CERN [1] aims to measure the lifetime of the 
$\pi^+\pi^-$ atom in its ground state with high precision. The atom 
predominantly decays into two neutral pions. 
The measurement of the quantity $\Gamma_{2\pi^0}$ will 
allow one to determine the difference $|a_0-a_2|$ 
of the strong $S$-wave $\pi\pi$ scattering lengths with isospin $I=0,2$ and 
to check the predictions for this quantity obtained in the standard version of 
ChPT [2], and to investigate the nature of spontaneous chiral symmetry
breaking in QCD [3].
In order to perform this programme, the theoretical 
expression for the width must of course be known with a precision that 
matches the accuracy of the lifetime measurement of DIRAC. 

We present a general 
expression for the corresponding decay width in the framework of QCD 
(including photons).
This expression  is proportional 
to the 
square of the relativistic $\pi^+\pi^-\to\pi^0\pi^0$ scatterting 
amplitude ${\cal A}$, contains all terms at 
leading and next-to-leading order in isospin breaking, 
and is valid to all orders in the chiral expansion [4]: 
\begin{eqnarray}
\Gamma_{2\pi^0}=\frac{2}{9}\,\alpha^3
p^\star{\cal A}^2(1+K),
\end{eqnarray}
where $p^\star=(M_{\pi^+}^2-M_{\pi^0}^2-\alpha^2 M_{\pi^+}^2/4)^{1/2}$. 
The quantities ${\cal A}$ and $K$ are expanded in powers of $\alpha$ 
and $m_d-m_u$. We count $\alpha$ and $(m_d-m_u)^2$ as small parameters of 
order $\delta$: 
\begin{eqnarray}
{\cal A}&=&a_0-a_2+h_1\,(m_d-m_u)^2+h_2\,\alpha + o(\delta),
\nonumber\\[2mm]
K&=&f_1\,(m_d-m_u)^2+f_2\,\alpha\ln\alpha+f_3\,\alpha + o(\delta),
\end{eqnarray}
where the scattering lengths $a_0$, $a_2$ and the coefficients $h_i$ and $f_i$ 
are evaluated in the isospin symmetry limit, i.e. at $m_u=m_d=\hat m$ and 
$M_\pi=M_{\pi^+} = 139.57$ MeV. 
The quantities $h_i$ and $f_i$ parameterize the corrections to the 
leading-order Deser's et al. type [5] formula. 
For the factor $K$ we get the exact expression (without any chiral 
expansion)
\begin{eqnarray}
K=\frac{\Delta_\pi}{9M_{\pi^+}^2}\,(a_0+2a_2)^2
+\frac{2\alpha}{3}\,(1-\ln\alpha)\,(2a_0+a_2)+o(\delta).
\end{eqnarray}
Therefore, the formula (1) relegates the problem of the calculation 
of the $\pi^+\pi^-$ atom lifetime to the evaluation of the physical 
on-mass-shell scattering amplitude for the process 
$\pi^+\pi^-\rightarrow\pi^0\pi^0$ to any desired order 
in the chiral expansion.  

Next we apply Chiral Perturbation Theory at one loop to analyse the 
formula for the $\pi^+\pi^-$ atom lifetime (1). 
We obtain the analytic expressions 
for the next-to-leading order $O(\alpha)$ and $O(m_d-m_u)^2$ corrections 
and pin them down numerically.

\bigskip

\noindent [1] 
B.~Adeva {\it et al.}, CERN proposal CERN/SPSLC 95-1 (1995). 

\noindent [2]
J.~Gasser and H.~Leutwyler, Phys. Lett.  125B (1983) 325; 
 J.~Bijnens, G.~Colangelo, G.~Ecker, J.~Gasser,  
and M.~E.~Sainio, Phys. Lett. B374 (1996) 210.

\noindent [3]
M.~Knecht, B.~Moussallam, J.~Stern, and N.~H.~Fuchs, 
Nucl. Phys. B 457 (1995) 513; ibid. B 471 (1996) 445. See also Girlanda's talk
at this meeting.

\noindent [4] 
A. Gall, J. Gasser, V.~E. Lyubovitskij, and A. Rusetsky, 
Phys. Lett. B 462 (1999) 335.

\noindent [5]
S.~Deser, M.~L.~Goldberger, K.~Baumann, and 
W.~Thirring,  Phys. Rev. {96} (1954) 774;
J.~L.~Uretsky and T.~R.~Palfrey, Jr., Phys. Rev. { 121} (1961) 1798;
S.~M.~Bilenky, Van Kheu Nguyen, L.~L.~Nemenov, and F.~G.~Tkebuchava,  
Yad. Fiz. { 10} (1969) 812 .

\newpage 

\setcounter{equation}{0}

\begin{center}
{\Large\bf Spectrum and decays of hadronic atoms}

\bigskip

J. Gasser$^1$, V.E. Lyubovitskij$^{2,3}$, 
 and \underline{A.G. Rusetsky}$^{1,2,4}$  

{$^1$ Institute for Theoretical Physics, University of Bern,
Sidlerstrasse 5, CH-3012, Bern, Switzerland}

{$^2$ Bogoliubov Laboratory of Theoretical Physics, Joint Institute
for Nuclear Research, 141980 Dubna, Russia}

{$^3$ Department of Physics, Tomsk State University, 634050 Tomsk, Russia}

{$^4$ HEPI, Tbilisi State University, 380086 Tbilisi, Georgia}
\end{center}

\bigskip

A general survey of the theoretical approaches to the description of hadronic
atom characteristics (spectrum and decay width) is given. We confront 
nonrelativistic effective Lagrangian approach, as well as other
field-theoretical approaches based on relativistic equations, with the 
potential scattering theory formalism for the hadronic atom problem.
This comparison is essential in the view of several ongoing or planned 
experiments (DIRAC at CERN, PSI, DEAR at LNF-INFN, KEK, Uppsala) that are
aimed at the high-precision measurement of hadronic atom characteristics.
These measurements will enable one to directly extract the strong 
scattering amplitudes from the data, and thus contribute a valuable piece of
information to our knowledge of the dynamics of strong interactions at low
energy.

The analysis of the data in the above-mentioned experiments are carried out
with the use of the relations given in Ref. [1]. Up to the corrections in
the next order in isospin breaking, these relate the energy-level
shift of the hadronic atom $\Delta E$ and its partial decay width into the
channel with neutral isotopic partners, to the certain isotopic combinations
of the strong scattering lengths. However, in order to fully exploit the 
high-precision data available from recent and future experiments on hadronic
atoms, one has to consistently evaluate the corrections to the relations given
in Ref.~[1]. Using the powerful technique based on the effective 
nonrelativistic Lagrangians, it is demonstrated that in the next-to-leading 
order in isospin breaking, and in all orders in chiral expansion, the partial
decay width of $\pi^+\pi^-$ atom into $2\pi^0$ is given by [2]
\begin{eqnarray}
\Gamma_{2\pi^0}&=&\frac{2}{9}\, \alpha^3\,(M_{\pi^+}^2-M_{\pi^0}^2
-\frac{1}{4}\,\alpha^2\,M_{\pi^+}^2)^{1/2}\,\,{\cal A}^2(1+K)
\nonumber\\[2mm]
K&=&\frac{\Delta_\pi}{9M_{\pi^+}^2}\,(a_0+2a_2)^2
+\frac{2\alpha}{3}\,(1-\ln\alpha)\,(2a_0+a_2)
\end{eqnarray}
and ${\cal A}$ stands for the regular part of {\it physical}  scattering 
amplitude for $\pi^+\pi^-\to \pi^0\pi^0$ at threshold, calculated at the 
leading order in isospin breaking, and in all orders in chiral expansion.
The strong energy-level shift of the $\pi^-p$ atom is also expressed solely in
terms of the regular part of {\it physical} $\pi^- p\to\pi^- p$ scattering 
amplitude at threshold, calculated at the leading order in isospin breaking.
The relations between the observables of hadronic atoms and the physical
scattering amplitudes obtained by using the nonrelativistic Lagrangian
technique, provide a desirable generalization of the relations from Ref. [1]
that now contains {\it all} leading-order isospin-breaking corrections.
These relations are obtained under a very general assumptions about the mass
differences of isotopic partners, and do not resort explicitly to the chiral
expansion of the corresponding strong amplitudes. In order to extract the
strong scattering lengths from the physical scattering amplitudes that are
actually measured in the experiment, one has, however, to use chiral expansion
for these amplitudes, and establish such relations in a given order in chiral
expansion. 

By using chiral expansion, it is demonstrated that the potential scattering
theory approach in its present form does not account for all isospin-breaking
corrections to the strong amplitude. The dominant isospin-breaking
contributions which are neglected in the potential approach, have been
explicitly identified. First of all, these are effects of direct quark-photon
coupling, that are encoded in the "electromagnetic" counterterms of the chiral
Lagrangian. Then, there are effects coming from "tuning" of the quark mass so
that the common mass of the pion triplet in the isospin-symmetric world
coincides, by definition, with the charged pion mass. We conclude, that for
comparison to the results of potential approach, it is necessary to match the
latter with ChPT in the isospin-broken phase, in order to consistently take
into account above effects.

\bigskip

\noindent [1] 
S.~Deser, M.~L.~Goldberger, K.~Baumann, and W.~Thirring,  
Phys. Rev. 96 (1954) 774.

\noindent [2] 
A. Gall, J. Gasser, V.~E. Lyubovitskij, and A. Rusetsky, 
Phys. Lett. B 462 (1999) 335.


\setcounter{equation}{0}

\begin{center}
{\Large\bf Calculation of the breakup
probability of $\pi^+\pi^-$ atom in a target with a high
accuracy}

\bigskip
\underline{L.Afanasyev}, M.Jabitski, A.Tarasov, and O.Voskresenskaya\\
Laboratory of Nuclear Problems \\
Joint Institute for Nuclear Research \\
141980 Dubna, Moscow Region \\
Russia\\

\end{center}

\bigskip

The interaction of $\pi^+\pi^-$-atom ($A_{2\pi}$) with matter is of great importance for
the DIRAC experiment [1] as the $A_{2\pi}$ breakup (ionization) in such
interactions is exploited to observe $A_{2\pi}$ and to measure its
lifetime. In the experiment the ratio of the number of $\pi^+\pi^-$-pair
from the $A_{2\pi}$ breakup inside a target to the number of produced
$A_{2\pi}$ (called the probability of $A_{2\pi}$ breakup) will be
measured. The measurement of the $A_{2\pi}$ lifetime is based on the
comparison of this experimental value with its calculated dependency
on the lifetime. So the accuracy of this calculation is
essential for the extraction of the lifetime.

In the DIRAC proposal the probability of $A_{2\pi}$ breakup was calculated
by evaluating of a set of differential equations for populations of
$A_{2\pi}$ states [2]. These equations describe the evolution of the
populations, when $A_{2\pi}$ traverses the target, for all atomic states
with the principal quantum number less than some limit.  The initial
condition for this set of equations is given by the probability of the
$A_{2\pi}$ production in various quantum states [3].  As a solution we
can get three probabilities for the case behind the target: 1) the
summed population $P_{\mathrm{dsc}}$ of all discrete states considered
in this set; 2) an estimate for the summed population of all other
discrete states $P_{\mathrm{tail}}$; 3) the probability of $A_{2\pi}$
annihilation $P_{\mathrm{anh}}$. The remainder is the probability of
the $A_{2\pi}$ breakup $P_{\mathrm{br}}$:
$$
P_{\mathrm{br}}=1 - P_{\mathrm{dsc}} - P_{\mathrm{tail}} -P_{\mathrm{anh}}\,.
$$

The accuracy of this procedure is mainly defined by the accuracy of
the cross sections for the $A_{2\pi}$ interaction and target atoms. Now
approaches are available to calculate these cross sections in the
Glauber approximation [4] which takes into account all multi-photon
exchanges instead of only the single-photon exchange in the first Born
approximation as used before. This could provide an accuracy in the
cross sections at the level less than 1\% and hence almost the same
accuracy for $P_{\mathrm{br}}$.

To estimate a limit in the accuracy of the $P_{\mathrm{br}}$ calculation 
based on probabilities we have repeated all our calculation of
ref.[2] using the parabolic basis for the $A_{2\pi}$ states description
instead of the spherical one as in [2]. In both cases we neglect 
interference effects in the $A_{2\pi}$ description but its contribution
should be different for another basis. So, the difference in
$P_{\mathrm{br}}$ obtained with two bases shows a principal limit
in accuracy for this method. In the following table the results of the
calculation are shown for the $A_{2\pi}$ momentum 4.7~GeV/$c$, the lifetime
$3.0\cdot10^{-15}$~s and target thicknesses  equivalent in multiple
scattering to 30~$\mu$m Ta.
\begin{center}
\begin{tabular}{|c|c|c|c|c|c|} \hline
   & $Z$ & $P^{\mathrm{sph}}_{\mathrm{br}}$
                & $P^{\mathrm{par}}_{\mathrm{br}}$
                        & $\Delta P_{\mathrm{br}}/P_{\mathrm{br}}$
                                 & $\Delta \tau / \tau$   \\ \hline
Al & 13 & 0.223 & 0.229 & 0.0292 & 0.079\\
Ti & 22 & 0.326 & 0.330 & 0.0125 & 0.030\\
Fe & 26 & 0.435 & 0.438 & 0.0069 & 0.018\\
Ni & 28 & 0.470 & 0.473 & 0.0059 & 0.017\\
Mo & 42 & 0.540 & 0.543 & 0.0044 & 0.015\\
Ta & 73 & 0.671 & 0.673 & 0.0026 & 0.015\\
Pt & 78 & 0.704 & 0.706 & 0.0022 & 0.017\\ \hline
\end{tabular}
\end{center}
The value of $\Delta \tau / \tau$ could be considered as the bias of the
method. To go lower than 1\% we need to consider the process of
the $A_{2\pi}$ passage through the target in terms of amplitudes instead of
probabilities.

\bigskip

\noindent [1] B. Adeva et al., {\it Lifetime measurement of $\pi^+\pi^-$   
atoms to test low energy QCD predictions}, (Proposal to the SPSLC, CERN/SPSLC
95--1, SPSLC/P 284, Geneva 1995).

\noindent [2] L.G. Afanasyev and A.V. Tarasov, {\it Yad. Fiz.} {\bf 59}
(1996) 2212.

\noindent [3] L.L. Nemenov,  {\em Yad.Fiz.} {\bf 41} (1985) 980.

\noindent [4] L. Afanasyev, A. Tarasov and O. Voskresenskaya {\it J.Phys.G} 
{\bf 25} (1999) B7--10.

\newpage 

\setcounter{equation}{0}

\begin{center}
{\Large\bf Pionium interacting with matter: \\ I. Formalism for coherent
interaction}

\bigskip

T. Heim$^1$, K. Hencken$^1$, \underline{D. Trautmann}$^1$, and G. Baur$^2$

${}^1$Institut f\"ur Physik, Universit\"at Basel, Switzerland \\
${}^2$Institut f\"ur Kernphysik, Forschungszentrum J\"ulich,
Germany
\end{center}

\bigskip

The experiment DIRAC, currently being performed at CERN [1], 
aims at measuring the
lifetime for pionium, i.e. $\pi^+\pi^-$-atoms,  in its ground state to better
than 10\% accuracy thus providing a crucial test for chiral perturbation
theory.
Attaining the required precision hinges on an accurate calculation of
all electromagnetic processes, competing with the strong interaction, to a
precision of a few percent. Therefore we have applied the semi-classical
formalism (SCA) to calculate the electromagnetic excitation and ionization,
i.e. breakup, of the pionium to such high precision.

The SCA has been established [2--4] as a powerful tool to
investigate such excitation processes even at relativistic energies.
In this framework we describe an arbitrary collision process in terms of a
trajectory associated with the projectile and characterized by an impact
parameter $b$. The amplitude for the transition---excitation or
ionization---of
the target system from an initial state $|i\rangle$ to a final
state $|f\rangle$ is given in first order of the interaction by the matrix
element
\begin{equation}
a_{fi}^{(1)}(b)=\frac{1}{{\rm i}\hbar}\int_{-\infty}^\infty{\rm d}t
\langle f|H_{\rm int}(\vec{R}_b(t))|i\rangle,
\end{equation}
where $\vec{R}_b(t)$ describes the projectile's trajectory specified by the
impact parameter in the target's rest frame (see figure).

\noindent
\begin{minipage}{70mm}
\setlength{\unitlength}{0.7pt}
\begin{center}
\begin{picture}(270,140)
\put(0,80){\vector(1,0){270}}
\put(0,30){\vector(1,0){270}}
\put(70,30){\vector(2,1){97}}
\put(70,0){\vector(0,1){130}}
\put(170,50){\vector(0,1){80}}
\linethickness{1pt}
\put(70,30){\vector(0,1){50}}
\put(170,80){\vector(1,0){25}}
\put(73,115){$x$}
\put(173,115){$x'$}
\put(272,22){$z$}
\put(272,72){$z'$}
\put(0,65){$t=-\infty$}
\put(220,65){$t=+\infty$}
\put(75,18){$O$}
\put(175,66){$O'$}
\put(73,64){$\vec{b}$}
\put(150,56){$\vec{R}_b$}
\put(190,85){$\vec{v}$}
\put(70,30){\circle*{14}}
\put(170,80){\circle*{5}}
\put(150,87){$Z_P$}
\put(51,37){$A$}
\end{picture}
\end{center}
\end{minipage}
\hfill
\parbox{95mm}{%
{\bf Figure:} Semi-classical picture of a projectile with charge
$Z_P$ moving on a trajectory with impact parameter $b$ past a
target atom $A$ consisting of particles of opposite charge.
$\vec{R}_b(t)$ describes the projectile's trajectory specified by the
impact parameter in the target's rest frame.
}

In our case, in the rest-frame of the pionium, the target material's ion can
be treated as a classical particle moving on a straight-line trajectory
$\vec{R}_b(t)=(b,0,\beta ct)$ at nearly the speed of light, while the
pionium at
the origin is treated quantum mechanically. As we are only interested in
pions forming atom-like complexes, their relative velocity must be small
(of order $v_\pi/c\approx\alpha$). Hence non-relativistic hydrogenic wave
functions are perfectly appropriate for the initial and final states
$|i\rangle$ and $|f\rangle$ of the pionium. On the other hand,
the complex charge distribution of the target atoms is taken into account
by including a screening function in the scalar potential experienced by the
pionium. In our calculations, we use screening functions that reproduce
exactly
expectation values of powers of the radial variable obtained with full
Dirac-Hartree-Fock-Slater wave functions for the heavy ion or atom [5, 6].
Therefore by neglecting magnetic terms (estimated to contribute no more
than 0.4\%),
the interaction Hamiltonian $H_{\rm int}$ reduces to the scalar
(Li\'enard-Wiechert-)
potential between the ion with charge $Z_P$ and the pionium. In the
pionium's rest frame we thus have:
\begin{equation}
\Phi (\vec{r}, t)   = {\displaystyle\frac{Z_Pe}{2 \pi^2} \sum^N_{k = 1} A_k
    \int \frac{\exp\left[{\rm i} \vec{s} \cdot \left( \vec{r}
    - \vec{R}_b (t) \right)\right] }{s^2 +\alpha_k^2 - ( \beta s_z )^2}
     \;{\rm d}^3 s},
\end{equation}
where the screening parameters $A_k$ and $\alpha_k$  are
taken from [6]. Performing the integration over coordinate
space implied in the matrix element (1) with the two pions
positioned at $\pm\vec{r}/2$ from the pionium's center-of-mass,
we then obtain the impact parameter dependent transition amplitude in
first order of the scalar interaction:
\begin{eqnarray}
a^{(1)}_{fi} (b) & = & \frac{2 Z_P\alpha}{{\rm i}\beta} \sqrt{4
 \pi (2\ell_f + 1) (2\ell_i + 1)}\;(-1)^{m_f}
 \sum_{\ell,m} {\rm i}^{\ell - m} \sqrt{2\ell + 1}
 \left( \begin{array}{ccc} \ell_f & \ell & \ell_i \\ 0 & 0 & 0 \end{array}
 \right) \nonumber \\ & \times &
 \left[ 1 - (-1)^{\ell}\right] \left( \begin{array}{ccc} \ell_f
 & \ell & \ell_i \\ -m_f & m & m_i \end{array} \right)
 \sum^N_{k = 1} A_k \int^\infty_{0}\!\! s \; {\rm d}s 
 \frac{B_{\ell m}(b, q_0,s)}{s^2 + \alpha_k^2 - (q_0\beta)^2} F^{\ell}_{fi}
 \left( \frac{s}{2}\right),
\end{eqnarray}
with
$q_0 = (E_f - E_i)/(\beta\hbar c)$, the straight-line trajectory factor
$B_{\ell m} (b, q_0, s)$ [7], and the radial form factors
$F^{\ell}_{fi} (k)$, depending on the bound state or continuum wave
functions of the pionium, as appropriate. These form factors can easily
be evaluated using standard methods as described thoroughly in [8]
and [9].

Integrating the squared amplitude over the impact parameter $b$
yields the inelastic cross section $\sigma^{(1)}_{fi}$ for the transition
between the states $|i\rangle$ and $|f\rangle$ and the total cross section
is then obtained by summing over bound and integrating over continuum
final states---or by using the completeness relation for the set of final
states:
\begin{eqnarray}
\sigma_{{\rm tot},i}^{(1)}  =
 \sum_f\llap{$\displaystyle\int\;$} \sigma_{fi}^{(1)}  =
 16 \pi \left( \frac{Z_P \alpha}{\beta} \right)^2
 \int^\infty_{q_0} s \; {\rm d}s \left[ 1 -F^0_{ii}(s) \right]
 \left[\sum^N_{k =1} \frac{A_k}{s^2 + \alpha_k^2 - (\beta q_0)^2}\right]^2.
\end{eqnarray}

The determination of transition probabilities and cross sections thus
reduces to the accurate and fast calculation of the radial form factors.

In [8] we have shown the suitability of our approach under quite
general conditions. We have calculated the total inelastic cross section
in dependence of the kinetic energy of the heavy ion (in the rest frame of the
pionium) for different target ions. Hereby we have shown that the total
cross section, divided by the dominating (but trivial) factor $1/\beta^2$
in (4), is essentially constant in the energy range
of interest to experiment DIRAC, i.e., between 2~GeV and 10~GeV.
Furthermore we have calculated cross sections for various transitions
and break-up processes also in the screened Coulomb field of a nucleus 
both analytically as well as numerically.

Our on-going developments now concentrate on including magnetic interaction
terms, as well as all higher-order perturbation contributions.

\bigskip

\noindent
[1] L.L.~Nemenov et al., {\it Proposal to the SPSLC: Lifetime measurement
of $\pi^{+}\pi^{-}$-atoms to test low-energy QCD predictions,}
CERN/SPSLC 95-1, SPSLC/P 284.

\noindent
[2] J. Bang and J.M. Hansteen, {\it K. Dan. Vidensk. Selsk. Mat.
Fys. Medd.} {\bf 31} (1959) 13.

\noindent
[3] L. Kocbach, {\it Z. Phys. A} {\bf 279} (1976) 233.

\noindent
[4] D. Trautmann and F. R\"osel, {\it Nucl. Instr. Meth.}
{\bf 169} (1980) 259.

\noindent
[5] G.~Moli\`ere, {\it Z. Naturforsch.} {\bf 2a} (1947) 133.

\noindent
[6] F.~Salvat, J.D.~Martinez, R.~Mayol and J.~Parellada, {\it Phys.
Rev. A} {\bf 36} (1987) 467.

\noindent
[7] P.A.~Amundsen and K.~Ashamar, {\it J. Phys. B} {\bf 14} (1981)
4047.

\noindent
[8] Z.~Halabuka, T.A.~Heim, K.~Hencken, D.~Trautmann, and
R.D.~Viollier, {\it Nucl. Phys.} B{\bf 554} (1999) 86.

\noindent
[9] D.~Trautmann, G.~Baur and F.~R\"osel, {\it J. Phys. B} {\bf 16}
(1983) 3005.

\newpage 

\setcounter{equation}{0}

\begin{center}
{\Large\bf Pionium interacting with matter: \\
II. Formalism for incoherent interaction}

\bigskip

Thomas Heim$^1$, \underline{Kai Hencken}$^1$, Dirk Trautmann$^1$,  
and Gerhard Baur$^2$

$^1$ Institut f\"ur theoretische Physik,
Universit\"at Basel, Klingelbergstrasse 82, 4056 Basel, Switzerland.

$^2$ Institut f\"ur Kernphysik (Theorie), Forschungszentrum J\"ulich,
52425 J\"ulich, Germany. 

\end{center}

\bigskip

The atomic structure was included in the previous calculation only via the 
(elastic) form factor  $F(q^2)$ [1]. Of course the  (electrons of the) 
atoms can  
be excited as well. A simple estimate is that these incoherent/inelastic
processes are of the order $1/Z$ compared to the elastic/coherent one; 
therefore, in order to achieve the desired 1\% accuracy needed for the 
experiment, a more precise calculation is needed. 

We derive the cross section for the incoherent contribution starting from 
PWBA (Plane Wave Born Approximation) [2]. The electromagnetic
transitions of the atoms and the Pionium are given through the
electromagnetic tensor  $W^{\mu\nu}$, which due to gauge invariance
and current conservation, only  depends on two scalar structure
functions $W^{(1)}(q^2, qP)$ and $W^{(2)}(q^2, qP)$. This is of course well
known, e.g., from electron scattering [3]. Taking into  account only the
scalar interaction and neglecting the magnetic terms, only  $W^{(2)}$ is
important and is proportional to the square of the charge
transition. One advantage of this approach is that even
though the relative motion of atom and Pionium is relativistic, the
structure functions $W$ can be calculated in the individual rest
frames and therefore nonrelativistic approximations can be made. 

We neglect recoil corrections and find the total cross section to be
$$
\sigma = \int d\omega \ d\Delta\  d^2q_\perp \frac{4\alpha^2}{\beta^2} 
\frac{W_{2\Pi}(\omega,q^2) W_{2A}(\Delta,q^2)}{\left(q^2\right)^2} 
$$
with $\omega$ the excitation energy of the Pionium (in its rest frame)
and $\Delta$ the one for the atom. The momentum square of the photon
is given by
$$
q^2 = - \left( \frac{\Delta^2}{\beta^2\gamma^2} + 
\frac{\omega^2}{\beta^2\gamma^2} +
\frac{2\omega\Delta}{\beta^2\gamma} + q_\perp^2
\right) = - \left( q_l^2 + q_\perp^2 \right)
$$
The relative Lorentzfactor is given by $\gamma$, $\beta=v/c$.

We discuss approximations made to find closed expressions for the
total cross section. In them the $\omega$ and $\Delta$ dependence in
$q^2$ are replaced by average values. We also find the cross section 
in the elastic case to be identical to the one from the SCA
calculation, see previous contribution.

To get the total incoherent contribution, we sum over all possible
excitations of the atom. This so-called inelastic scattering function
$S_{inc}$ is then given by
$$
S_{inc}(k) = \int  d\Delta W_2(\Delta,k^2)
= \sum_{X\not=0} \left|F_{X0}(k)\right|^2 
$$
with $F_{X0}$ the transition form factor. For the calculation 
we make use of mean-field wave functions calculated within a
self consistent Dirac-Hartree-Fock-Slater calculation [4]. In terms of
the single electron wave functions $S_{inc}$ is then given by
$$
S_{inc}(\vec k) = Z - \sum_{j=1}^Z\sum_{l=1}^Z 
|\langle\Phi_j|\exp({\rm i}\vec{k} \cdot\vec{r})|\Phi_l\rangle |^2.
$$

The often used ``no correlation limit'' for $S_{inc}$ is found to be
not accurate enough. This is due to the fact that at the small $q^2$,
important in our case, atoms are mainly excited and not ionized;
therefore Pauli blocking, which is neglected in the no-correlation
limit, is important.

Results of calculations within this formalism are presented in the
next contribution.

\bigskip
\noindent [1] Z. Halabuka, T. A. Heim, K. Hencken, D. Trautmann, and R. D.
Viollier, Nucl. Phys. B554 (1999) 86.

\noindent [2] T. A. Heim, K. Hencken, D. Trautmann, G. Baur, in
preparation (1999). 

\noindent [3] T. deForest and J. D. Walecka, Adv. Phys. 15 (1966) 1.

\noindent [4] F. Salvat, J. D. Martinez, R. Mayol, and J. Parellada,
Phys. Rev. A 36 (1987) 467.

\newpage 

\setcounter{equation}{0}

\begin{center}
{\Large\bf Pionium interacting with matter: III. Results}

\bigskip

\underline{T. Heim}$^1$, K. Hencken$^1$, D. Trautmann$^1$, and G. Baur$^2$ 

${}^1$Institut f\"ur Physik, Universit\"at Basel, Switzerland \\
${}^2$Institut f\"ur Kernphysik, Forschungszentrum J\"ulich, 
Germany
\end{center}

\bigskip

We present results of a detailed investigation of the target-elastic and
target-inelastic electromagnetic cross sections for pionium scattering off
various target materials.
Within the closure approximation [1] the total electromagnetic 
cross sections from a given pionium initial state $i$ can be written as 

\bigskip
\hfill $\sigma_{\rm tot} = 16\pi\displaystyle\left(\frac{\alpha}{\beta}
\right)^2\int_0^\infty{\rm d}q_{\perp}\,q_{\perp}\frac{\Phi_A(k)}{q^4}
(1-f_{ii}^0(s))$ \hfill (1) 
\bigskip

\noindent
where the momentum transfer variables $s$ and $k$ refer to the rest frame of
the pionium and the atom, respectively.
$f_{ii}^0$ denotes the monopole formfactor for the pionium in the state $i$. 
For the target-elastic total cross section the atomic structure is
given by $\Phi_A(k)=(Z-F_{00}(k))^2$, where $F_{00}$ denotes the coherent 
formfactor of the electronic ground state orbitals only, as the contribution 
from the (point-like) nucleus has been separated out. 
We calculate the atomic formfactor in the framework of the 
Dirac-Hartree-Fock-Slater model, i.e., using  orbitals obtained from the 
numerical 
solution of the Dirac (or Schr\"odinger) equation for each occupied orbital.
However, the coherent formfactor is very well approximated by simple analytical 
expressions as given by Moli\`ere or Salvat {\em et al.} [2].

In the case of the target-inelastic total cross section, the atomic structure
is contained in $\Phi_A(k)=S_{\rm inc}(k)$  as given in the preceding 
contribution. We determine $S_{\rm inc}$ 
using the same orbitals as for the coherent formfactor.
In a simple ``no-correlation limit'', $S_{\rm inc}(k)=Z-[F_{00}(k)]^2/Z$. 
This approximation is {\em not} sufficiently accurate, as it
completely neglects Pauli blocking and thus over-estimates the cross section.
Our approach with 
Dirac-Hartree-Fock orbitals correctly accounts for Pauli blocking in
$S_{\rm inc}(k)$. Moreover it 
affords determining the contributions to the pionium cross sections for
each atomic shell individually. This information cannot be extracted from
tabulated values of formfactors and scattering functions [3] where only 
the combined contributions of all atomic shells are given. 

The diagram on the right shows the integrand of (1), multiplied with the
integration variable $q_\perp$ (thus representing more clearly the relative
magnitude of the individual contributions to the integral on a logarithmic
scale for $q_\perp$). 

\noindent
\begin{minipage}[b]{82mm}
The target material is Ti ($Z=22$), pionium is initially in its ground 
state. Note that the choice of atomic excitation energy $\Delta$ affects
the integrand only at $q_\perp$ smaller than the range of dominating 
contributions. The main contributions to the integral come from a complex 
interplay of photon propagator, atomic, and pionium structure at 
$1\le q_\perp\le 100$~a.u.
Note also that the outer shells contribute much more to the target-inelastic
cross section than the inner shells. Not only are there more electrons in 
outer shells than in inner shells, but 
the contribution of each individual target electron is approximately
proportional to its principal quantum number. 
Finally, note that the target-inelastic cross
section is significantly larger than the target-elastic divided by $Z$.  
Thus neither the ``no-correlation'' limit, nor the simple scaling
approximation $\sigma_{\rm inc}\approx\sigma_{\rm coh}/Z$ are accurate
enough.
\end{minipage}
\hfill
\begin{minipage}[b]{85mm}
\epsfxsize 85mm
\leavevmode\epsffile{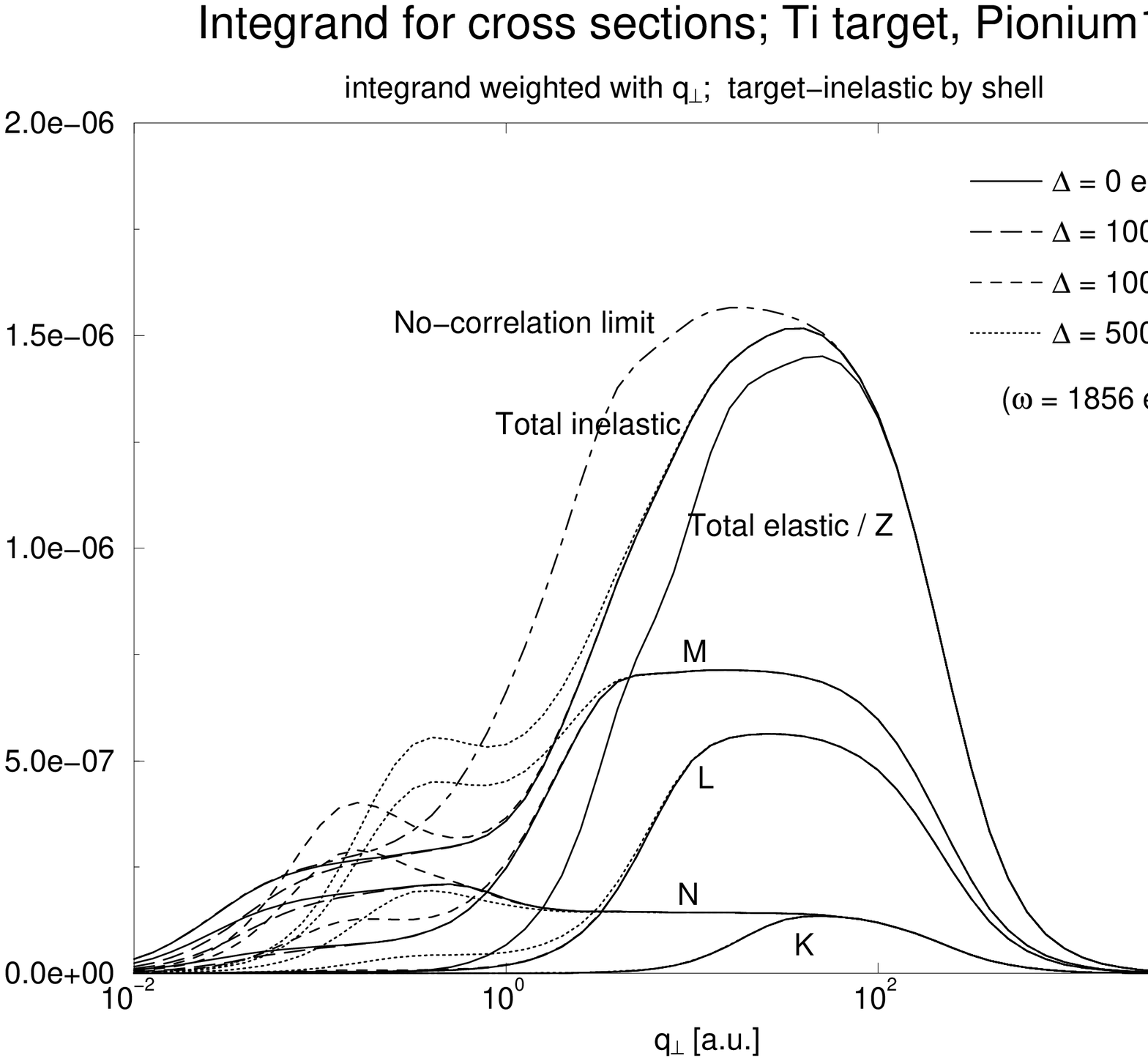}
\end{minipage}

\bigskip

\noindent [1] Z.~Halabuka, T.A.~Heim, K.~Hencken, D.~Trautmann, and
R.D.~Viollier, {\it Nucl. Phys.} B{\bf 554} (1999) 86.

\noindent [2] F.~Salvat et al., {\it Phys. Rev.} A{\bf 36}
(1987) 467; 
G.~Moli\`ere, {\it Z. Naturforsch.} {\bf 2a} (1947) 133.

\noindent [3] J.H.~Hubbell et al., {\it J. Phys. Chem. Ref. Data}
{\bf 4} (1975) 471; {\it ibid.} {\bf 6} (1977) 615;
J.H.~Hubbell and I.~\O verb\o, {\it ibid.} {\bf 8} (1979) 69.

\newpage 

\setcounter{equation}{0}

\begin{center}
{\Large\bf Another exotic relativistic atom: antihydrogen}

\bigskip

\underline{Gerhard Baur}$^1$, Kai Hencken$^2$, Helmar Meier$^2$, 
and Dirk Trautmann$^2$

$^1$ Institut f\"ur Kernphysik (Theorie), Forschungszentrum J\"ulich,
52425 J\"ulich, Germany. 

$^2$ Institut f\"ur theoretische Physik,
Universit\"at Basel, Klingelbergstrasse 82, 4056 Basel, Switzerland.

\end{center}

\bigskip

It is certainly difficult to do experiments with fast neutral
atoms. Recently, antihydrogen has been produced and detected at
LEAR/CERN [1] and Fermilab [2]. Cross-sections for the process

\bigskip
\hfill
        $Z + \overline p \rightarrow Z + \overline{H_0} + e^-$
\hfill $(1)$
\bigskip

are calculated [3], see also figure below. The formalism is very
similar to the one for calculating the breakup of Pionium in matter, 
see also these miniproceedings. Lamb-shift measurements
of hydrogen [4] and antihydrogen in flight, which are planned at
Fermilab are briefly discussed, see Refs. [4,5,6].

Maybe related experimental techniques could also be relevant for the
measurement of the Lamb-shift in Pionium.  Of course, the properties
of these atoms have very different scales (lifetime, size, mass,
\dots). This 2s-2p energy splitting is an interesting quantity since
the scattering lengths $a_0$ and $a_2$ are involved in a different
combination as compared to the one, which enters in the Pionium
lifetime [7]. 

\begin{center}
\epsfxsize 7cm
\leavevmode\centerline{\epsffile{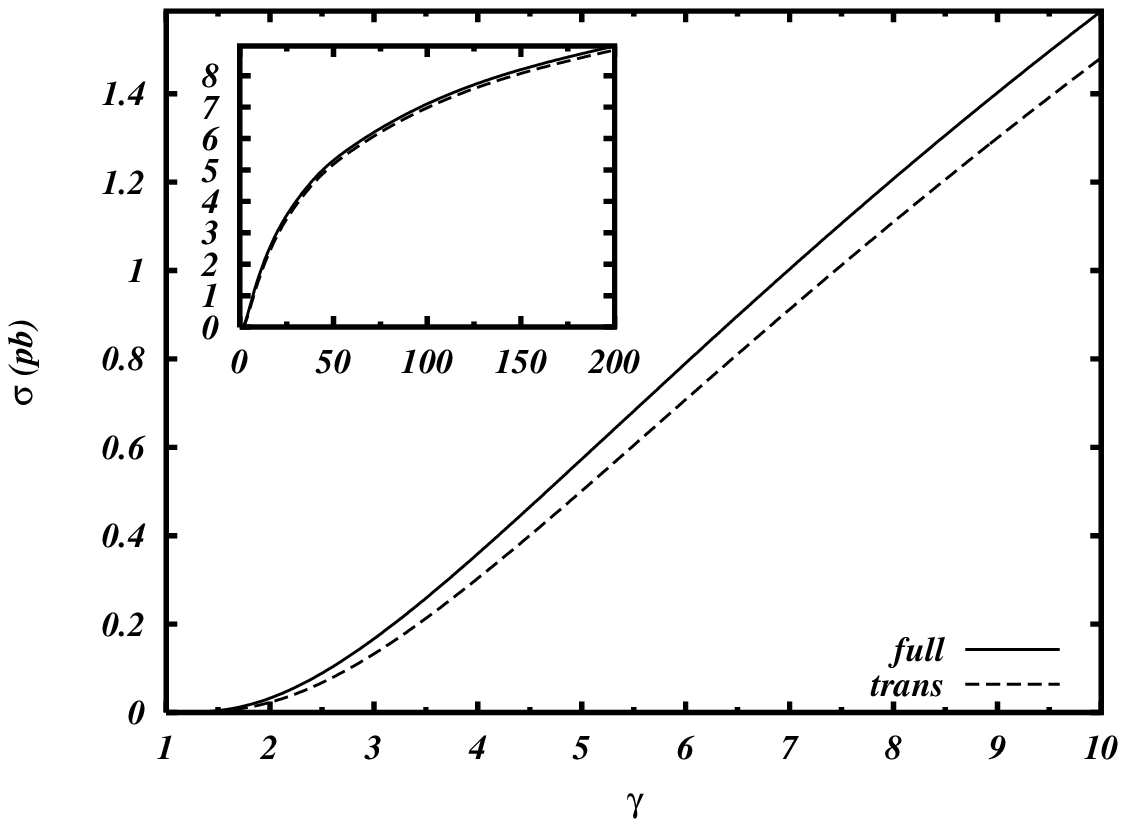}}
\\
\end{center}
Fig.~1: Cross sections for the reaction eq.~1 for $Z=1$. The Lorentz factor 
$\gamma$ of 
the incoming antiproton in the Fermilab experiment [2] was in the range of 
5.6--6.7, corresponding to an antiproton momentum range of 5200--6200 MeV$/c$.
A cross section of $\sigma_{1s+2s}=(1.12 \pm 0.14 \pm 0.09)$pb was found. (The
cross section $\sigma_{ns}$ for capture into the $s$-state with principal
quantum number $n$ scales as $1/n^3$.) 

\bigskip

\noindent [1] G. Baur et al., Phys. Lett. {\bf B368} (1996) 251.

\noindent [2] G. Blanford et al., Phys. Rev. Lett. {\bf 80} (1998) 3037.

\noindent [3] H. Meier, Z. Halabuka, K. Hencken, D. Trautmann, and
G. Baur, Eur. Phys. J. {\bf C5} (1998) 287.

\noindent [4] V. V. Parkhomchuk, Hyp. Int. {\bf 44} (1988) 315.

\noindent [5] C. T. Munger, S. J. Brodsky, I. Schmidt, Phys. Rev. {\bf
D49} (1994) 3228.

\noindent [6] G. Blanford et al., Phys. Rev. {\bf D57} (1998) 6649.

\noindent [7] L. L. Nemenov et al., CERN/SPSLC 95-1, SPSLC/P 284,
1999.

\newpage 

\setcounter{equation}{0}

\begin{center}
{\Large\bf Multiphoton exchange in pionium-matter interaction}

\bigskip

\underline{J. Cugnon}$^1$ and D. Vautherin$^2$

$^1$Physics Department B5, University of Li\`{e}ge, B-4000 Sart Tilman
Li\`{e}ge 1, Belgium\\
$^2$LPTPE, Universit\'{e} Pierre et Marie Curie, 4, Place Jussieu, case 127
F-75252 PARIS CEDEX 05, France
\end{center}

\bigskip

The interaction of a  high velocity pionium atom colliding with an ordinary
atom
 is investigated
by studying the evolution of the internal wave function driven by the
time-dependent Coulomb field generated by the ordinary atom. This evolution is
governed by the
following time-dependent equation:
\begin{equation}
\frac{\partial\psi}{\partial t} = - \frac{i}{\hbar} \left(H_0 +
V\left(t\right)\right) \psi,
\end{equation}
where $H_0$ is the intrinsic pionium Hamiltonian and where $V(t)$ is the
Coulomb interaction between the pionium and the external atom. Using the
expansion of $\psi$ in
terms of the eigenstates
$\psi_i$ of
$H_0$, i.e.
\begin{equation}
\psi = \sum\limits_{i}^{} c_i e^{- \frac{i}{\hbar}\varepsilon_i t} \psi_i,
\end{equation}
one obtains a set of coupled linear differential equations for the coefficients
$c_i$:
\begin{equation}
\frac{dc_i}{dt} = - \frac{i}{\hbar} \sum^{}_j V_{ij}(t) c_j (t).
\end{equation}
\par
Here, we solve this set of equations in a subset of bound pionium states. We
assume that the atom is moving along a straight line with a constant velocity,
in the rest frame of the pionium. We adopt the dipole approximation, i.e.  that
 the Coulomb interaction is reduced to the interaction between the electric
dipole of the pionium and the electric field generated by the atom. For the
latter, the Moli\`{e}re form factor is used. Equations (3) are integrated with
proper initial conditions $c_j (t=0)$ = $\delta_{ij}$, corresponding to the
pionium being in state $\psi_i$ initially. The cross section for the
excitation
 from
state $\psi_i$ to state $\psi_j$ is obtained by integrating over the impact
 parameter $b$
\begin{equation}
\sigma_{i \rightarrow j} = \int^{\infty}_0 db\ b\ |c_j (t \rightarrow
\infty)|^2.
\end{equation}
\par
Keeping the initial values $c_j = \delta_{ij}$ in the r.h.s. of the
equations (3) yields the usual Born
approximation[1]. We also show
that  expression (4) with Born approximation is equivalent to the
field theoretical expression for
one photon exchange, as obtained in ref.[2], when recoil correction is
 neglected, provided a one-to-one correspondance is introduced between the
impact parameter and the tranverse momentum transfer. This allowed us also to
correct the dipole approximation used in $ V(t)$ at small impact parameter
(details can be found in ref.[3]).
The difference between the exact solution of Eq. (3) and the Born approximation
can thus be interpreted as due to multiphoton exchanges.
\par
 This contribution has been evaluated numerically for a few illustrative cases.
 The
dependance upon the impact parameter, the atomic number of the external
atom and  the incident energy is investigated. For the most intense transition
1S $\rightarrow$ 2P, the effect of multiphoton exchange is a reduction of
the transition
cross section. This reduction is effective for reduced impact parameter
$\tilde{b}$ ($=b/a$, $a$ being the pionium Bohr radius) lying between
$\sim$ 0.5
and $\sim$ 2. Let us mention that the bulk of the cross section involves
$\tilde{b}$ between  $\sim$ 0.5 and $\sim$ 5 and that the dipole approximation
is only effective for $\tilde{b}$ less than unity.

\bigskip

\noindent [1] J.-L. Basdevant, "M\'{e}canique Quantique", Ellipses, 1986

\noindent [2] S. Mr\'{o}wczy\`{n}ski, Phys. Rev. {\bf D36} (1987) 1529

\noindent [3] J. Cugnon and D. Vautherin, Proc. of "Hadronic Atoms and
Positronium in
the Standard Model", ed. by M.A. Ivanov {\it et  al}, 1999, p.128

\newpage 

\setcounter{equation}{0}

\begin{center}

{\Large\bf The DEAR experiment at DA$\Phi$NE}

\bigskip

Carlo Guaraldo

(on behalf of the DEAR collaboration)

INFN-Laboratori Nazionali di Frascati, C.P.13, 
Via E.Fermi 40, I-00044 Frascati, Italy

\end{center}

\bigskip

A new era in the field of low energy kaon physics will begin 
with the start of the
new DA$\Phi$NE
low-energy, high-luminosity 
$\rm e^{+}$$\rm e^{-}$~collider.
The objective of the 
{\bf DEAR} (
{\bf D}A$\Phi$NE {\bf E}xotic {\bf A}tom {\bf R}esearch) experiment
is the determination of the isospin dependent $\rm \bar{K}N$ scattering
lengths via the 
measurement of the strong interaction shifts and widths of the 
kaonic hydrogen and kaonic deuterium {\it K}-series lines [1].
In practice, in the case of kaonic hydrogen, the $\rm K_{\alpha}$ line
at $\sim$~6.15~keV
is the most important one for the determination of the shift and width
of the {\it 1s} level.

The challenging goal DEAR is aiming at is to measure the 
K$_{\alpha}$ line shift in hydrogen
to a precision of 1~$\%$, the width to a precision of a few percent,
and kaonic deuterium for the first time.

The main challenge of DEAR is to detect a very weak
x-ray signal out of the background of an
$\rm e^{+}$$\rm e^{-}$~collider with high statistics and high precision. 
Only recently these X rays have been observed
for the first time (KpX at KEK [2]) with an overall 
statistics of only 114 counts in the $\rm K_{\alpha}$ peak.

To overcome the background problem 
Charge-Coupled Devices (CCDs) 
are used as x-ray detectors.
CCDs are characterized by a good energy resolution and
by an unprecedented background rejection capability [3].

A kaonic atom is formed when a negative kaon enters a target, 
loses its kinetic energy through ionization and excitation of the
molecules of the medium and eventually is captured,
replacing the electron, in an excited orbit. 
Various cascade processes deexcite
this kaonic atom to the ground state.

When a kaon reaches low-$n$ states with small angular
momentum, it is absorbed through a strong interaction with the nucleus.
This strong interaction causes a shift in the energies of the low-lying levels
from their purely electromagnetic values, whilst the finite lifetime 
of the state turns out in an increase in the observed level width.

The shift $\epsilon$ and the width $\Gamma$ of the {\it 1s} state 
of kaonic hydrogen are related in a fairly model-independent way
to the real and imaginary part of the complex $s$-wave scattering
length, $a_{K^-p}$:
$ \epsilon + \frac{i}{2} \Gamma = 2 \alpha^3\mu^2 a_{K^-p} = 
({\rm 412\,eV fm^{-1}}) \cdot a_{K^-p} $
(Deser-Trueman formula [4]),
where $\alpha$ is the fine structure constant and 
$\mu$ the reduced mass of the $K^{-}p$ system.
A similar relation applies to the case of kaonic deuterium and the 
corresponding scattering length, $a_{K^-d}$.

These observable scattering lengths are related to the isospin dependent
scattering lengths $a_{0}$ and $a_{1}$ in the following way:
$a_{K^-p} = \frac{1}{2} (a_0 + a_1)~~,~~~
a_{K^-d} = \frac{1}{2} \left(\frac{m_N+m_K}{m_N+m_K/2}\right) 
(a_{0}+3a_{1})~+~C~.$
In the case of deuterium, the first term represents the lowest-order
impulse approximation, in which the kaon scatters on each ``free''
nucleon. The second term, $C$, contains all
high-order contributions, including three-body effects. 
This term can be larger than the first term,
thus the extraction of the two isospin scattering lengths
from an observation of the kaonic hydrogen and kaonic deuterium
scattering lengths requires a dedicated analysis.
An accurate determination of the $K^{-}N$ isospin dependent
scattering lengths will place strong constraints
on low energy  $K^{-}N$ dynamics, which in turn constraints
the SU(3) description of chiral symmetry breaking.

Crucial information about the nature of chiral symmetry breaking,
and to what extent chiral symmetry is broken,
is provided by the calculation of the meson-nucleon sigma terms.
The meson-nucleon sigma terms are defined as
the expectation value of a double commutator of the
chiral symmetry breaking part of the strong-interaction
Hamiltonian.
The low energy theorem
relates the sigma terms to the meson-nucleon
scattering amplitude.
A phenomenological procedure starting from the experimental amplitudes
is then used to determine the sigma terms and therefore measure
chiral symmetry breaking.
According to an evaluation based on the uncertainties 
in the phenomenological procedure
the sigma terms can be extracted at the level of 20~$\%$,
by combining the precision measurement to be performed by
DEAR with the bulk of most recent analyses of low energy $K^{\pm}N$
scattering data.
The sigma terms are also important inputs for the determination
of the strangeness content of the proton.

Presently, the DEAR experiment is installed on DA$\Phi$NE
and ready to start data taking.

\vspace*{2mm}

[1] S.~Bianco et al., LNF-Preprint, LNF-98/039 (P) (1998); 
Rivista del Nuovo Cimento, November 1999.

[2] T.M.~Ito et al., 
Phys. Rev. A 58 (1998) 2366.
\hfill

[3] J.-P.~Egger et al., Part. World 3 (1993) 139.
\hfill

[4] S. Deser {\em et al.}, Phys. Rev. {\bf 96}
(1954) 774;
T.L. Truemann, Nucl. Phys. {\bf 26} (1961) 57.
\hfill

\newpage 

\setcounter{equation}{0}

\begin{center}
{\Large\bf Pionic hydrogen: status and outlook.}

\bigskip

Leopold M. Simons (for experiment R-98-01.1 at PSI)

Paul Scherrer Institute, 5232 Villigen PSI, Switzerland
\end{center}

\bigskip

The measurement of the strong interaction shift and width of the ground state
 of the pionic hydrogen atom allows to determine two different 
 linear combinations 
of the two isospin separated 
s--wave scattering lengths of the pion nucleon system. Past experiments 
 at PSI measured the 3--1 transition energy in pionic hydrogen 
 and deuterium [1,2,3]. The measurement combined a highly efficient 
 stopping arrangement (cyclotron trap) and a sophisticated cylindrically 
 bent Bragg crystal set-up with newly developed CCD detectors. 
The shift was finally determined with an accuracy 
 of better than $10^{-2}$. The error in the determination of the 
 width of the ground state
 level in hydrogen 
 is a factor of almost one  order of magnitude  worse than the error in the 
shift value, which inhibits 
 a determination of both  isospin separated scattering lengths on the percent
 level from the pionic hydrogen measurement alone. The situation can be 
 improved in a combined analysis 
 with the pionic deuterium shift measurement, 
but then one has to rely on a theoretical  understanding 
of the pionic deuterium system.

A new proposal at PSI aims at a direct  determination
of the    width 
 of the ground state of pionic hydrogen on the level of $10^{-2}$ [4]. In 
order to achieve this 
the experiment should meet several requirements as are  a higher number of 
detected X--rays and a thorough  understanding of the response function of the 
 crystal spectrometer. Most importantly it should also be able to disentagle 
the strong interaction broadening from other line broadening effects as 
the Doppler effect. Pionic hydrogen atoms may gain kinetic energy
 by converting binding energy difference into kinetic energy, a process 
known as Coulomb deexcitation.
 A recent experimental investigation of this effect took place at PSI.
 The distortion of neutron time of flight spectra from the charge
 exchange reaction has been measured  in liquid hydrogen and in gas at 
 a pressure of 40 bar (room temperature) [5]. The measured spectra clearly
 show the effect of different transition steps converted into kinetic energy.

The statistics problem was solved by stopping more pions in a newly built 
 cyclotron trap  and by increasing 
 the efficiency of the crystal spectrometer by using spherically bent 
 crystals as well as bigger CCD detectors. 
In this way a factor of  more than one order of magnitude was gained
 in count rate. In addition the peak to background ration as well as the 
 resolution of the spectrometer had been improved. 
First experiments with pionic deuterium show the 
 high quality of the data achieved by these measures[6].

For a better determination of the response function an ECR (Electron Cyclotron
 Resonance) source is presently being set up which  produces X--rays  from 
 hydrogen--like ions with high intensity. Their line widths are expected 
to be an order of magnitude 
 narrower than the expected resolution of the spectrometer. Under these 
 circumstances a detailed measurement of the response function and even 
 an optimization of the crystals will be feasible.

The third problem will be attacked in two steps. 
First the 2--1 and the 3--1 transitions 
 in pionic hydrogen will be measured at about three different pressures. The 
 influence of the Doppler effect on the line shape will be different for 
 the different transitions and pressures whereas the strong interaction 
 width will remain the same. Monte Carlo simulations have shown that 
 a suitable fitting procedure will allow to reach an accuracy in the 
 width determination of $ 3\times 10^{-2}$.

 In a second step a simultaneous mesurement of muonic 
and pionic hydrogen X--rays 
 will be performed. The muonic hydrogen X--rays are affected by Doppler 
 effect in  a similar way as their pionic counterparts. In contrast to them
 they are not broadened by strong interaction,however, and 
therefore offer a good chance to determine 
 the influence of the Doppler effect [7].   A further check of the theory 
 of the cascade processes  will be carried out by measuring the velocity 
 of the system at the instant of  the charge exchange process by the 
 neutron time of flight method. The pionic X--ray measurements will start 
 end of year 2000 and first results can be expected in the year after.

\bigskip   

\noindent [1] D. Sigg et al., Nucl. Phys. A 609 (1996) 269.

\noindent [2] D. Chatellard et al., Nucl. Phys. A625 (1997) 855.

\noindent [3] H.-Ch. Schr\"{o}der et al., to be published in Phys. Lett. B

\noindent [4] Proposal R-98-01.1 at PSI.

\noindent [5] A. Badertscher et al., Phys. Lett. B392 (1997) 278. 

\noindent [6] P. Hauser et al. Phys. Rev. C58 (1998) R1869. 

\noindent [7] V. E. Markushin, private communication.

\newpage 

\setcounter{equation}{0}

\begin{center}
{\large\bf Pion--kaon scattering}\\[0.5cm]
Ulf-G. Mei\ss ner\\
FZ J\"ulich, IKP(Th), D-52425 J\"ulich, Germany\\[0.3cm]
\end{center}

\noindent Pion--kaon scattering has been considered in refs.[1,2] in the
framework of three flavor chiral perturbation theory to one loop accuracy.
To that order, all low energy constants have been determined from other
reactions (and invoking large $N_c$ arguments), so that phase shifts, threshold
and subthreshold parameters can be predicted. Despite the rather large
threshold energy, $\sqrt{s}_{\rm thr} =M_K +M_\pi \approx 633\,$MeV, the
one loop corrections to the S--wave scattering lengths are about 25\%. The
uncertainties due to the various parameters at next--to--leading order
are also discussed in~[1]. Other approaches and extensions (explicit
resonances~[3], heavy kaon approach~[4], the inverse amplitude method~[5]
extended to coupled channels~[6]) all lead to very similar threshold 
parameters. The experimental values scatter over large ranges, see e.g.
the figure in~[2]. Also, two groups have performed a dispersion theoretical
analysis~[7,8]. While $a_0^{3/2}$ and  $a_1^{1/2}$ agree with the chiral
predictions, the dispersion relation result for  $a_0^{1/2}$ is sizeably 
larger. A new dispersive analysis including also the SLAC data from 1988
is in progress~[9]. In that context, we have also derived a sum rule,
\begin{equation}
\frac{1}{F_K F_\pi} = \frac{4}{\pi} \int_{(M_\pi+M_K)^2}^\infty ds'
\frac{{\rm Im}~F^- (s',t)}{(s'-s)(s'-u)}~.
\end{equation}
Even if one assumes that Im~$F^-$ is entirely given by the $K^* (892)$
and approximating the latter by a $\delta$--function, one obtains $F_K F_\pi
\approx (127\,$MeV$)^2$ not far off the empirical value of $(102\,$MeV$)^2$.
Other work in progress contains a one loop analysis of $\pi^- K^+ \to \pi^0
K^0 \gamma$~[10], which is needed to calculate the width of $\pi K$ atoms.
Finally, an analysis of $\pi K$ scattering in GCHPT is also available~[11]. 

\section*{References}
\noindent [1] V. Bernard, N. Kaiser and Ulf-G. Mei\ss ner,
Nucl. Phys. B357 (1991) 128.\\
\noindent [2] V. Bernard, N. Kaiser and Ulf-G. Mei\ss ner,
Phys. Rev. D43 (1991) R2757.\\
\noindent [3] V. Bernard, N. Kaiser and Ulf-G. Mei\ss ner,
Nucl. Phys. B364 (1991) 283.\\
\noindent [4] A. Roessl, Nucl. Phys. B555 (1999) 507.\\
\noindent [5] A. Dobado and J. Pelaez, Phys. Rev. D56 (1997) 3057.\\
\noindent [6] J.A. Oller et al., Phys. Rev. D59 (1999) 074001.\\
\noindent [7] C.B. Lang, Fortschr. Physik 26 (1978) 509.\\
\noindent [8] N. Johannesson and G. Nilsson, Nuov. Cim. 43A (1978) 376.\\
\noindent [9] B. Ananthanarayan, P. B\"uttiker and  Ulf-G. Mei\ss ner, 
forthcoming.\\
\noindent [10] B. Kubis and  Ulf-G. Mei\ss ner, forthcoming.\\
\noindent [11] M. Knecht et al., Phys. Lett. B313 (1993) 229.

\newpage 

\setcounter{equation}{0}

\begin{center}
{\Large\bf $K_{l4}$ decays at DA$\Phi$NE}

\bigskip

Patrizia de Simone 

for the KLOE Collaboration

Laboratori Nazionali di Frascati dell'INFN, Frascati, Italy

\end{center}

\bigskip

The capabilities of the KLOE detector [1], and the over constrained 
kinematic of the $\phi$'s decays at DA$\Phi$NE [2] will give the
opportunity to improve our knowledge on the $K_{l4}$ decays.

The experimental results obtained so far are dominated by the work of
L. Rosselet et al. [3] which measures the ~$\pi^+\pi^-$~ phase shifts
with $\simeq 30000$ $K_{e4}^+$ events collected.
The phase shift difference $(\delta^o_o -\delta^1_1)$ was 
determined in five bins of the invariant mass $s_{\pi}$ of the dipion
system, and the related isoscalar S-wave scattering length was~
$a_o^o = 0.26 \pm 0.05$. This result must be compared with the CHPT
prediction ~$a_o^o = 0.20 \pm 0.01$ [4]. 
The apparent discrepancy between the CHPT prediction and the 
L. Rosselet experiment may be interpreted as a  manifestation of an
unusual low value of the quark-antiquark condensate: a new 
measurement of the isoscalar S-wave scattering length $a_o^o$
with a substantially smaller error can determine if this
discrepancy is statistically significant. 

Two methods have been proposed to measure the 
isoscalar S-wave scattering length $a_o^o$: the Pais Tre\-i\-man 
method [5], and the Maximum Likelihood Method ($MLM$) [6].
The $MLM$ requires to do approxi\-ma\-tions/assum\-ptions on the form
factors, while the Pais Treiman method is model independent.
The two methods are complementary and both will be considered at KLOE.
Moreover, since the Pais Treiman method was devised for a detector
with a uniform efficiency over the whole phase space such as
we expect KLOE to be, this will be the first time that 
this method will be applied (a variant of the Pais Treiman 
method has been used by L. Rosselet et al. [3]).

The statistical errors on the $K_{e4}^{\pm}$ parameters has been evaluated
using the $MLM$ method [6], in particular, concerning the 
~$\pi\pi$~ phase shifts, the error on the
isoscalar S-wave scattering length has been estimated using the
J.L. Basdevant et al. parametrisation [7]. For $300000$ $K_{e4}^{\pm}$
events the estimated error is ~$\delta a^o_o = 0.01$. This estimate
is purely statistical and apply to a ``perfect'' detector, i.e., 
one which covers the whole phase space with unity efficiency everywhere.
This is close to being true for KLOE, that it is a hermetic detector operating
at~ DA$\Phi$NE producing self-tagging $K^{\pm}$ pairs, with high
reconstruction efficiency of neutral and charged 
low energy particles that will be
controlled at the level of $\simeq 10^{-3} \div 10^{-4}$.

In one year of running at~ DA$\Phi$NE at the target luminosity of
${\cal L}_o = 5\times10^{32}$ cm$^{-2}$s$^{-1}$, given that the cross
section for ~$e^+e^- \rightarrow \phi$~ at the $\phi$ resonance peak
is $\simeq 3.8 \mu$b,  
the number of ~$\phi$ decays into charged kaons will be:
\[
   N(\phi \rightarrow K^+K^-) \simeq 9.4 \times 10^{9}
\]
and the number of $K_{e4}$ decays are:
\[
   N(K_{e4}^{\pm}) =
         \underbrace{1.88\times10^{10}}_{N(K^{\pm})}\times
         \underbrace{3.9\times10^{-5}}_{Br(K_{e4}^{\pm})}\times
         \underbrace{.60}_{\epsilon_{K^{\pm}}}\times
         \underbrace{.95}_{\epsilon_{minv}^{cut}}
         \simeq~ 4.2\times10^5
\]

\noindent
including the efficiency $\epsilon_{K^{\pm}}$ 
to select the $\phi$ decays in charged
kaon events, and the cuts ($\epsilon_{minv}^{cut}$) 
on the Dalitz invariant mass plot.  
To collect $300000$ $K_{e4}^{\pm}$ decays in one year,
the reconstruction efficiency of the
$K_{e4}^{\pm}$ events has to be $\simeq 71$\%.

The KLOE detector at the Frascati $\phi$-factory DA$\Phi$NE
is fully operational and has recently started to collect its
first data. During a short test run period 220 nb$^{-1}$ of
integrated luminosity has been accumulated and has been used 
to study the detector performance and to fully test the 
reconstruction program. The detector is ready to collect and
reconstruct data at the expected luminosity. 

\bigskip

\noindent [1] J. Lee-Franzini, The second DA$\Phi$NE Physics Handbook,
              ed. L. Maiani et al. (Frascati, 1995) p.761.

\noindent [2] G. Vignola, Proc. Workshop on Physics and Detectors
              for DA$\Phi$NE, ed. G. Pancheri et al. (Frascati, 1991) p.11.

\noindent [3] L. Rosselet {\it et al.}, Phys. Rev. D15 (1977) p.574.

\noindent [4] J. Gasser and H. Leutwyler, Phys. Lett. B125 (1983) p.321.

\noindent [5] A. Pais and S.B. Treiman, Phys. Rev. 168 (1968) p.1858.

\noindent [6] M. Baillargeon and P.J. Franzini, The second 
              DA$\Phi$NE Physics Handbook, ed. L. Maiani et al. 
              (Frascati, 1995) p.413.
 
\noindent [7] J.L. Basdevant {\it et al.}, Nucl. Phys. B72 (1974) p.413.

\newpage 

\setcounter{equation}{0}

\begin{center}
{\Large\bf A new measurement of the $K^+\rightarrow \pi^+\pi^- e^+ \nu$ 
decay.}

\bigskip

Stefan Pislak for the E865 collaboration \\

Yale University
\end{center}

\bigskip

\begin{figure}[htb]
\begin{center}
\includegraphics[angle=0,width=0.49\linewidth]{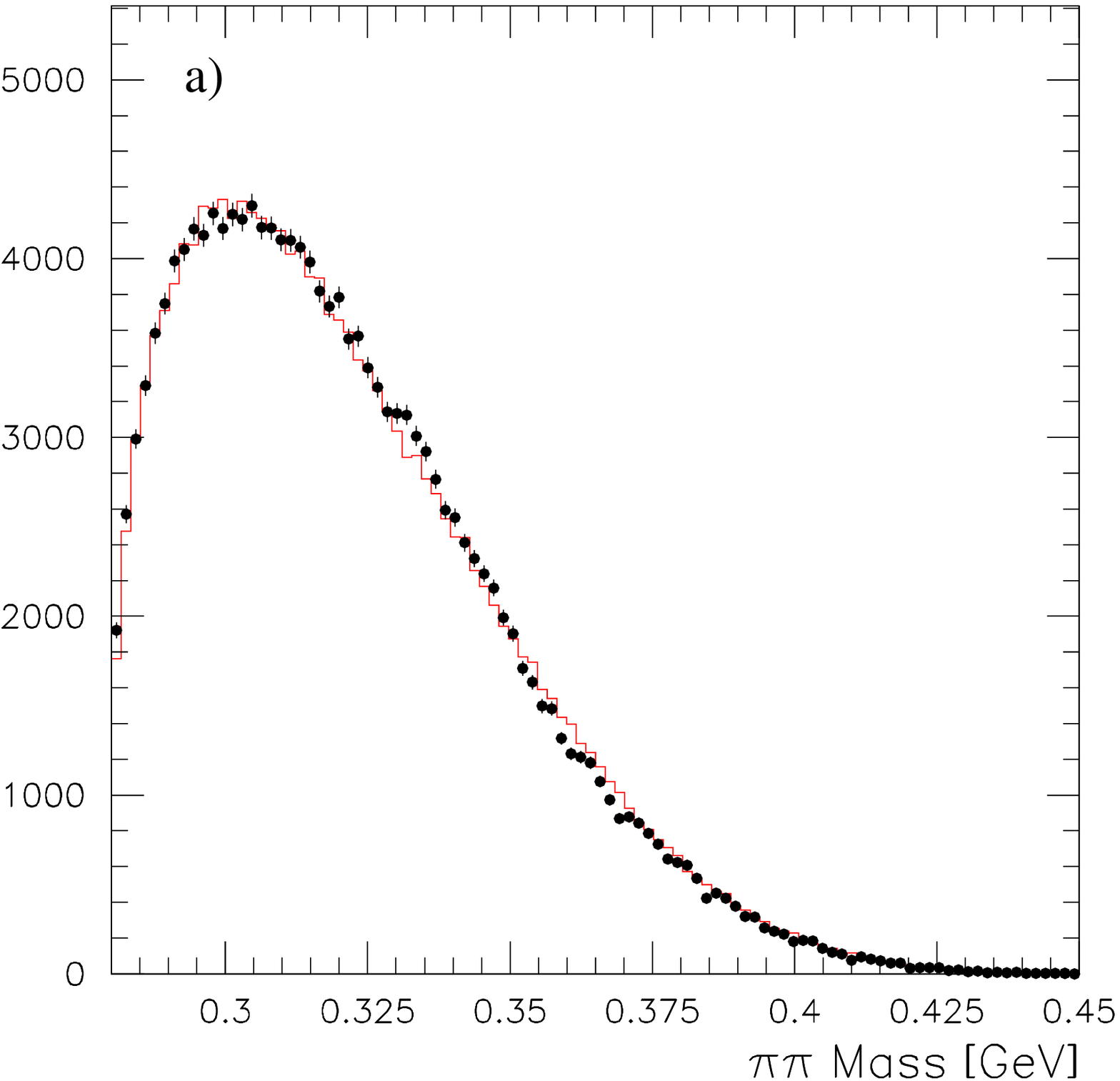}
\includegraphics[angle=0,width=0.49\linewidth]{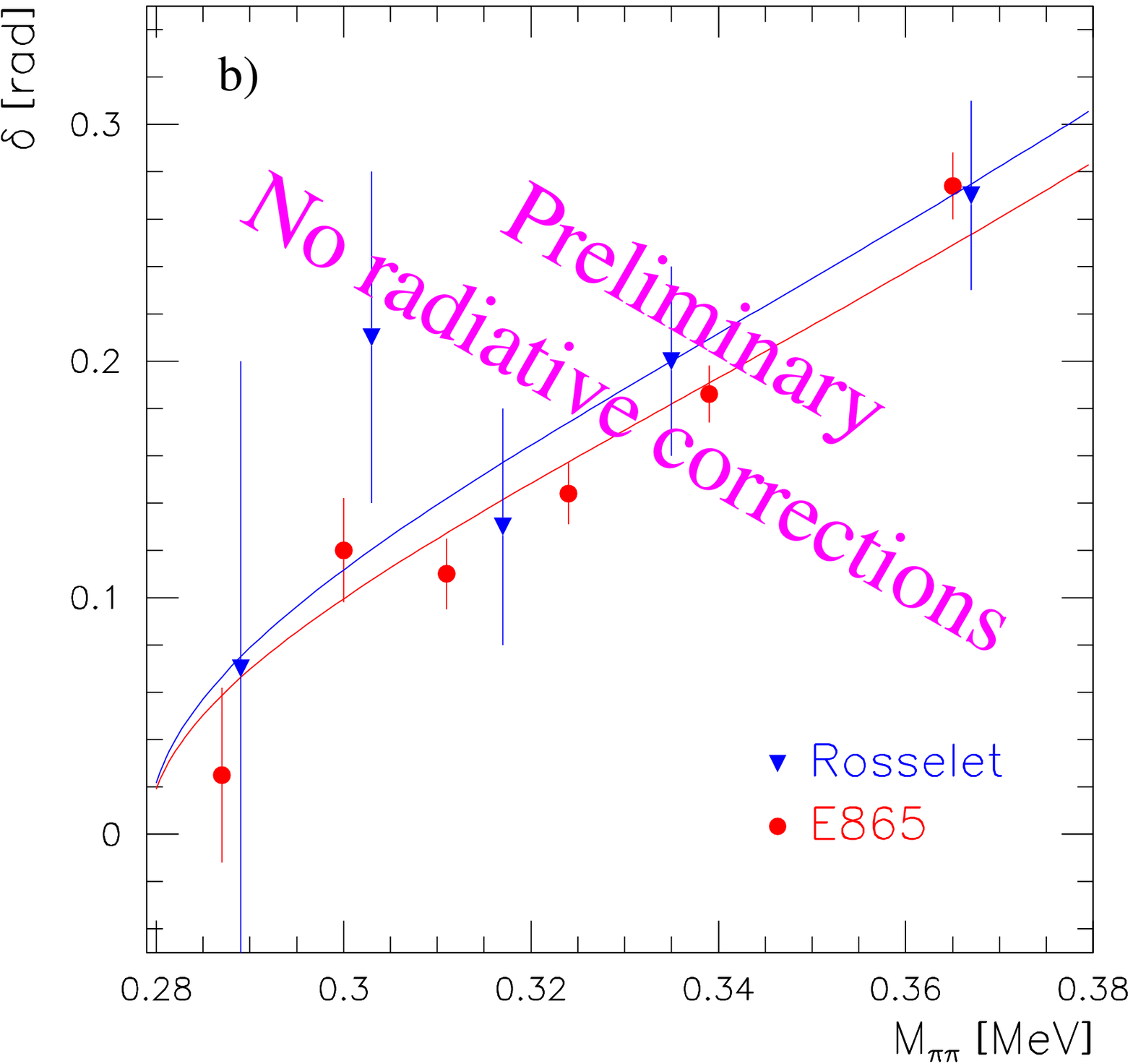}
\end{center}
\caption[]{a) Distribution of the reconstructed $\pi\pi$ mass. The dots
represent the data, and the histogram shows a Monte Carlo simulation.
b) Dependence of the phase shift difference 
$\delta\equiv\delta^0_0-\delta^1_1$ on the $\pi\pi$ mass. The fits
are based on the work by Basdevant~\emph{ et al.} [5].} 
\label{fig:data}
\end{figure}

The measurement 
was performed by experiment E865 at the Brookhaven Alternating Gradient 
Synchrotron employing an apparatus that has been described in previous 
publications [1]. The selected $K_{e4}$ candidates were required to 
have an unambiguous identification of the $\pi^-$ to suppress
background originating from $K^+$ decays with a $\pi^0$ 
(mostly $K_{\pi2}$, $K_{e3}$, and $K_{\mu3}$), giving an $e^+e^-$ pair.
In order to reject the $K^+\rightarrow\pi^+\pi^+\pi^-$ ($K_{\tau}$) 
background, we require an unambiguous identification of the $e^+$ and a
kaon, reconstructed from the three charged decay products, which
does not track back to the target.
The basis of the rejection of accidental background is the requirement
to have three trajectories consistent with having come from a common 
vertex, and a timing spread between the trajectories consistent with 
the resolution. Our final signal sample contains 437,000 $K_{e4}$ 
candidates including 2\% background events.
These data represent a more than tenfold increase in
statistics compared with previous experiments~[2]. 
Fig.~\ref{fig:data}a) shows the reconstructed $\pi\pi$
mass in comparison with the Monte Carlo simulation, one of 
many control plots, demonstrating our good understanding of the
data.

These data were then used to fit phase shift differences and form
factors as a function of the $\pi\pi$ mass. In a first analysis, a
formalism as it was described by Rosselet~[2] was employed. Radiative 
corrections were not included but their calculation is in 
progress~[3]. A preliminary fit of the phase shift differences as
a function of the $\pi\pi$ mass is shown in Fig.~\ref{fig:data}b). 
An estimation of the phase shifts and form factors based on the 
parameterization proposed by Amor\'{o}s and Bijnens [4] and the 
determination of the branching ratio is in progress.

\bigskip

\noindent [1] R.~Appel~\emph{ et al.}, hep-ex/9907045, to be published
in Phys. Rev. Let. {\bf 83}, 29 November 1999.

\noindent [2] L.~Rosselet~\emph{ et al.}, Phys. Rev. {\bf D15} (1977) 574.

\noindent [3] M.~Knecht, these proceedings.

\noindent [4] G.~Amor\'{o}s and J.~Bijnens, J.Phys. {\bf G25} (1999) 1607.

\noindent [5] J.~L.~Basdevant~\emph{ et al.}, Nucl. Phys. {\bf B72} (1974)
413.

\newpage 

\setcounter{equation}{0}

\begin{center}
{\Large\bf $K_{\ell 4}$ decays: a theoretical discussion}

\bigskip

G.Amor\'os

Lund University
\end{center}

\bigskip

To measure the s-wave of the $\pi \pi$ scattering, the experiments need to focus
on processes where the inelastic channels are supressed. The decay of 
the pionium system and the $K_{\ell 4}$ decays are competitive and 
independent ways. 
This talk will be related with $K_{\ell 4}$ decays where the supposed precision 
for the 
experiment will force the calculation to higher orders in CHPT [1]. 
An overview of 
the status of the theoretical calculation and the influence on the experimental 
parametrization [2] is given. 
A first part of the talk is related with a general and model independent 
parametrization for the $K_{\ell 4}$ decays. Previous study of the prediction 
for the relevant form factors in several models and in the framework of CHPT 
allows us to consider the linear dependence with the kinematical variables as 
the natural one. This parametrization could decrease the number of variables 
to fit. 
In the second part of the talk some results about the two-loops calculation 
is given. Decay constants and masses for the complete $SU(3)$ octect are 
obtained.

\bigskip

\noindent [1] G. Amor\'os, J. Bijnens and P. Talavera, work in preparation.

\noindent [2] G. Amor\'os and J. Bijnens, J. Phys. G. 25 (1999) 1607.

\newpage 

\setcounter{equation}{0}

\begin{center}
{\Large\bf Chiral phase transitions at zero temperature}
\bigskip

Jan Stern

I.P.N., Groupe de Physique Theorique,  Universite de Paris-Sud,
F91406 Orsay Cedex , France
\end{center}

\bigskip

It is argued that in QCD , quark loops progressively suppress order
parameters of chiral symmetry breaking as the number $N_f$ of light flavours
increases. This could result into a non-trivial chiral phase structure of
the theory considered as a function of $N_f$ and $N_c$. For $N_f$ close to
the first critical point $n_{crit}(N_c)$, one may expect [1] a reduction of the
size of the chiral condensate $<\bar qq>$ and an enhancement of its
fluctuations leading to an important violation of the Zweig rule (i.e. to a
breakdown of large$N_c$ predictions) in the scalar iso-scalar channel.
Starting from observed correlation between $\bar uu$ and $\bar ss$, it is
possible to estimate the fall-off of the condensate $<\bar uu>_{N_f}$ as the
number of light spectators varies from $N_f=2$ to $N_f=3$ [2]. Existing data
are consistent with a strong $N_f$ - dependence : The three- flavour
condensate could be suppressed by as much as a factor 2 relative to the
two-flavor condensate,  possibly indicating that
the real world ($N_{f}=2-3 ,N_{c}=3$)  could already feel the influence of
a nearby phase transition [2]. It is important to make a careful distinction
between $<\bar uu>_{2}$ and $<\bar uu>_{3}$ in all phenomenological analysis.
The two-flavour condensate (in appropriate units) is measurable in
precise low-energy $\pi-\pi$ scattering experiments and is strongly correlated
with the quark mass ratio $r=2 m_s/(m_{u}+m_{d})$ [1]. The main goal of these
experiments remains the determination of the two-flavour condensate,
in order to examine to which extent the expected decrease
of the condensate $<\bar uu>_{N_f}$ towards the critical
point already starts to be visible at $N_f=2$. Whatever the experimental
answer to this question will be, additional experimental information will be
necessary to pin down the three-flavour condensate and to settle the
theoretical issue of a nearby phase transition as a common explanation of the
observed large Zweig rule violation in the vacuum-channel and of the $N_f$
dependence of the chiral symmetry breaking.  
This might require new experiments involving Kaons ,in which
the strange quark would play a more direct role than that of         
a ``sea-side-spectator''.

\bigskip

\noindent [1] S. Descotes, L. Girlanda, J. Stern: ''Paramagnetic Effect of
Light Quark Loops on Chiral Symmetry Breaking'', hep-ph/9910537.

\noindent [2] B. Moussallam: ''$N_f$  Dependence of the Quark Condensate from a
Chiral Sum Rule'', hep-ph/9909292.

\newpage 

\setcounter{equation}{0}

\begin{center}
{\Large\bf The two-flavor chiral condensate from low-energy $\pi\pi$
phase-shifts}

\bigskip

Luca Girlanda

I.P.N., Groupe de Physique Th\'eorique, Universit\'e de Paris-Sud~XI, F-91406
Orsay Cedex, France
\end{center}

\bigskip
Low-energy $\pi\pi$ scattering has been identified as the most promising
physical observable for measuring the quark condensate [1]. What can actually be
measured in $\pi\pi$ experiments is the two-flavor quark condensate, i.e.
$\langle \bar q q \rangle$ in the SU(2)$\times$SU(2) chiral limit [2].
Due to the properties of analyticity, unitarity and crossing symmetry and to
the fact that Goldstone bosons interact weakly at low energy, the amplitude
$A(s|,t,u)$ can be written, up to and including the two-loop level, in terms
of six parameters $(\alpha, \beta,
\lambda_1,\lambda_2,\lambda_3,\lambda_4)$. This low-energy representation
holds independently of the  size of $\langle \bar q q \rangle$. The explicit
expression can be found in Ref.~[3]. $\alpha$ and $\beta$ represents
essentially the amplitude and its slope at the symmetrical point $s=t=u=4/3
M_{\pi}^2$. They can be directly related to the two-flavor quark condensate
measured by the Gell-Mann--Oakes--Renner ratio $x_2=-2 \hat m \lim_{\hat m
\to 0} \langle \bar q q \rangle /(F_{\pi}^2 M_{\pi}^2)$.
In addition to constraints from chiral symmetry, the amplitude satisfies
also a set of crossing-symmetrical dispersion relations,
known as Roy Equations. Using as input experimental
phase-shifts in the 
medium- and high-energy region, they allow to determine the low-energy
amplitude in terms of two subtraction constants, usually identified with the
two scalar scattering lengths $a_0^0$ and $a_0^2$.
In fact the two subtraction constants can be related to each other, imposing
the matching of the amplitude to the experimental $P$-wave, which is well
known down to rather low energies, thanks to the $\rho$ resonance.
This results in the so-called Morgan-Shaw universal curve, that we quote in
the form given by Petersen in 1979,
\[
2 a_0^0 - 5 a_0^2 = 0.692 \pm 0.027  + 0.9 \left( a_0^0 - 0.3 \right)  + 1.2
\left( a_0^0 - 0.3\right)^2.
\]
By comparing the two low-energy representations one is able, in principle,
to determine 4 out of the 6 parameters of the chiral amplitude. A similar
program has been undertaken in Ref.~[3], in order to fix the parameters
$\lambda_1,\ldots ,\lambda_4$. It turns out that the outcome of this
analysis is only barely dependent on $\alpha$ and $\beta$, and hence on the
quark condensate. The slight $\alpha$- and $\beta$-dependence can be
conveniently approximated by linear functions
$\lambda_i^{\mathrm{exp}}(\alpha, \beta)$,
\[ 
\begin{array}{ll}
\lambda_1^{\mathrm{exp}} = \left[ -6.08 - 0.37 (\alpha -2) + 6.8 (\beta
-1.08) \right] \cdot 10^{-3}, & \lambda_2^{\mathrm{exp}} = \left[ 9.56 +
0.22 (\alpha -2 ) + 2.2 (\beta -1.08) \right] \cdot 10^{-3},\\
\lambda_3^{\mathrm{exp}} = \left[ 2.20 - 0.01 (\alpha -2) + 1.1 (\beta
-1.08) \right] \cdot 10^{-4}, & \lambda_4^{\mathrm{exp}} = \left[ -1.46 +
0.02 (\alpha -2 ) - 1.6 (\beta -1.08) \right] \cdot 10^{-4},\\
\end{array}
\]
while the experimental uncertainties on the $\lambda_i$ are independent of
$\alpha$ and $\beta$, $\delta \lambda_1^{\mathrm{exp}} = 2.2 \cdot 10^{-3}$,
$\delta \lambda_2^{\mathrm{exp}} = 0.5 \cdot 10^{-3}$,
$\delta \lambda_3^{\mathrm{exp}} = 0.6 \cdot 10^{-4}$,
$\delta \lambda_4^{\mathrm{exp}} = 0.12 \cdot 10^{-4}$.
With the parameters $\lambda_1,\ldots ,\lambda_4$ determined from the Roy
dispersion relations, the low-energy $\pi\pi$ observables are effectively
parametrized only by $\alpha$ and $\beta$, which can be fitted to the
experimental phase-shifts. However, in order to fully exploit the analytical
properties of the amplitude contained in the two-loop six-parametrical
formula of Ref.~[3], we can treat all the six parameters as fit variables,
and add  the $\lambda_i^{\mathrm{exp}}(\alpha,\beta)$ as additional
experimental points to the $\chi^2$. Moreover one can insert the constraint
of the Morgan-Shaw universal curve, which, using the two-loop expressions for
the scattering lengths~[3], can also be expressed in terms of the fit
variables ($\alpha,\beta,\lambda_1,\ldots ,\lambda_4)$.
Thus we define the $\chi^2$ as
\[
\chi^2 = \sum_n \left[ \frac{\delta_n (\alpha,\beta,\lambda_1,\lambda_2,
\lambda_3, \lambda_4) - \delta_n^{\mathrm{exp}}}{\sigma_n^{\mathrm{exp}}}
\right]^2 + \sum_{i=1}^4 \left[ \frac{\lambda_i -
\lambda_i^{\mathrm{exp}}(\alpha,\beta)}{\delta \lambda_i^{\mathrm{exp}}}
\right]^2 +
\chi^2_{\mathrm{Morgan-Shaw}}(\alpha,\beta,\lambda_1,\lambda_2,\lambda_3,
\lambda_4),
\]
where $\delta_n^{\mathrm{exp}}$ are the experimental phase-shifts, with
experimental error $\sigma_n^{\mathrm{exp}}$. It is straightforward to
include to the $\chi^2$ the combination $a_0^0 -  a_0^2$, once the pionium
lifetime will be measured.
Applying this fit procedure to the old $K_{e4}$ Rosselet data
[$\delta_n^{\mathrm{exp}} = \delta_0^0(E_n) - \delta_1^1(E_n)$], we find for
the $S$-wave isoscalar scattering length, $a_0^0=0.268 \pm 0.043$, and for the
combination relevant to the pionium lifetime, $a_0^0 - a_0^2 = 0.294 \pm
0.033$. Notice that the error on the difference of scattering lengths is
smaller
than the individual error on $a_0^0$, reflecting the importance of
properly taking into account all the correlations among the fit variables.
Using the relationship between $x_2$ and $(\alpha,\beta)$, known at one-loop
level, we get $x_2 = 0.71 \pm 0.13_{\mathrm{exp}} \pm 0.05_{\mathrm{theo}}$,
where the experimental and theoretical errors can be considered as mutually
uncorrelated. 

\bigskip

\noindent [1] N.H.~Fuchs, H.~Sazdjian and J.~Stern,
Phys. Lett. {\bf B 269} (1991) 183;
Phys. Rev. {\bf D 47} (1993) 3814.

\noindent [2] S.~Descotes, L.~Girlanda and J.~Stern, hep-ph/9910537.

\noindent [3] M.~Knecht, B.~Moussallam, J.~Stern and N.H.~Fuchs,
Nucl. Phys. {\bf B 457} (1995) 513;
 Nucl. Phys. {\bf B 471} (1996) 445.

\newpage 

\setcounter{equation}{0}

\begin{center}
{\Large\bf Radiative corrections to semi-leptonic decays}

\bigskip

Marc Knecht

Centre de Physique Th\'eorique, CNRS Luminy, Case 907, 
F-13288 Marseille Cedex 9, France
\end{center}

\bigskip

Semi-leptonic decays are an important source of information on hadronic 
matrix elements of the Standard Model vector and axial currents and, in the 
case of the $K^+\to \pi^+\pi^-e^+\nu_e$ decay, on low-energy $\pi-\pi$ 
scattering.
In order to meet the accuracy of two-loop calculations and the expected 
improvements on the experimental side in the near future (BNL-E865, KLOE, 
DIRAC,...), electromagnetic corrections to these processes, or isospin 
breaking contributions in general, have to be estimated in a reliable way.

The first step towards this aim has been proposed in Ref. [1]. In order to 
address the question of electromagnetic corrections in semi-leptonic 
processes on a systematical basis, the particle spectrum of chiral 
perturbation theory with virtual photons has been enlarged by including also 
the light leptons. The presence of radiative corrections affects the 
semi-leptonic amplitudes in two ways. First, the factorization property 
between the hadronic and the leptonic weak currents is lost, due to loop 
graphs involving both virtual photons and virtual leptons. Second, these 
additional loop graphs generate in general new divergences, which are taken 
care of neither by the Gasser-Leutwyler low-energy constants $L_i$ [2], nor by 
Urech's counterterms $K_i$ [3]. The structure of 
these divergences have been analysed at one loop, using 
super-heat-kernel methods [4], and a complete list of the corresponding 
counterterms has been given in [1]. As an illustration of this framework,
an explicit calculation of the decay rates for $\pi\to\ell\nu(\gamma)$ and 
$K\to\ell\nu(\gamma)$ has also been provided.
The extensions to the other semi-leptonic processes, and to the pion 
$\beta -$ decay, are in progress. 

A numerical estimate of the corresponding 
decay rates eventually requires a 
determination of the counterterms involved in each amplitude. 
This may be achieved following methods originally suggested in Ref. [5], 
combined with large-$N_C$ techniques [6]. 
From this point of view, the recent study [7] of the decay of pseudoscalar 
mesons into pairs of charged leptons, which, as far as some aspects of this question are concerned, may be 
considered as a ``weak SU(2) rotation'' of the $\pi\to\ell\nu$ decay, offers 
interesting perspectives.

\bigskip

\noindent [1] M. Knecht, H. Neufeld, H. Rupertsberger and P. talavera, 
{\it Chiral Perturbation Theory with Virtual Photons and Leptons}, 
hep-ph/9909284, to appear in Eur. Phys. J. {\bf C}.

\noindent [2] J. Gasser and H. Leutwyler, Nucl. Phys. {\bf B250}, 465 (1985).

\noindent [3] R. Urech, Nucl. Phys. {\bf B4333}, 234 (1995).\\
 ~~H. Neufeld and H. Rupertsberger, Z. Phys. {\bf C68}, 91 (1995); 
ibid. Z. Phys. {\bf C71}, 131 (1996).

\noindent [4] H. Neufeld, J. Gasser and G. Ecker, Phys. Lett. {\bf B438}, 106 91998).\\ 
~~H. Neufeld, Eur. Phys. J. {\bf C7}, 355 (1999).

\noindent [5] A. Sirlin, Rev. Mod. Phys. {\bf 50}, 573 (1978).

\noindent [6] G. 't Hooft, Nucl. Phys. {\bf B72}, 461 (1974).\\ 
~~E. Witten,  Nucl. Phys. {\bf B160}, 57 (1979).

\noindent [7] M. Knecht, S. Peris, M. Perrottet and E. de Rafael, 
{\it Decay of Pseudoscalars into Lepton Pairs and Large-$N_C$ QCD}, 
hep-ph/9908247, to appear in Phys. Rev. Lett..

\newpage 

\begin{center}
{\Large\bf Numerical solutions of Roy equations}

\bigskip

Gilberto Colangelo \\
Institut f\"ur Theoretische Physik
der Universit\"at Z\"urich \\
Winterthurerstr. 190, 8057 Z\"urich
\end{center}

\bigskip 

We solve numerically the Roy equations [1] for the $S$ and $P$ waves of the
$\pi \pi$ scattering amplitude in the low--energy region (below 800 MeV)
[2]. The imaginary parts above 800 MeV or of higher waves are taken either
from experiments (where available) or from theoretical modelling.  We are
able to solve the equations for any $S$ wave $I=0$ scattering length
$a_0^0$. At fixed $a_0^0$ solutions can be found only for $a_0^2$ (the $S$
wave $I=2$ scattering length) within a certain range, called the Universal
Band.

We then confront the solutions we find for various values of $a_0^0$ and
$a_0^2$ to the available experimental data. The corresponding $\chi^2$
analysis yields ellipses of given confidence level inside the Universal
Band. We confirm old analyses [3] of the $K_{e4}$ data [4] leading to an
$a_0^0$ between 0.20 and 0.30 [5]. This machinery is ready to be used with
forthcoming $K_{e4}$ data which should reduce drastically the error on
$a_0^0$.  

\bigskip

\noindent [1] S.M. Roy, {\it Phys. Lett.} 36B (1971) 353.

\noindent [2] B. Ananthanarayan, G. Colangelo, J. Gasser, H. Leutwyler an
G. Wanders, work in progress.

\noindent [3] J. L. Basdevant, C. D. Froggatt and J. L. Petersen, {\it
  Nucl. Phys.} B72 (1974) 413.

\noindent [4] L.\ Rosselet et al., {\it Phys. Rev.} D15 (1977) 574.

\noindent [5] M.M.\ Nagels et al., {\it Nucl. Phys.} B147 (1979) 189.

\newpage

\setcounter{equation}{0}
\setcounter{footnote}{0}

\begin{center}

{\Large\bf Analysis of bound state problems with non relativistic Lagrangians}

\bigskip

V. Antonelli\\

{Dipart. di Fisica, Universit\`a di Milano e Milano-Bicocca e INFN 
Sez. Milano, Via Celoria 16, Milano, Italia}\\ 
\end{center}

\bigskip

We discuss the use of Non Relativistic Lagrangians for the study of bound 
state characteristics and in particular of hadronic atoms. This study can 
give important information about strong interactions at low energies. Hence 
it has been the main aim of many experiments performed in the last years and 
of others, like DIRAC at CERN and DEAR at DA$\Phi$NE, which are running now 
or planned to start soon. The nature of the problem (and mainly the fact that 
the bound state components momenta are much smaller than their masses) 
suggests the use of non relativistic lagrangians. 
This approach, adopted in a number of recent 
publications [1-2], has proved to be probably the most efficient way to study 
these systems in quantum field theories.
In this connection, we feel that it is useful to discuss general properties of
the non relativistic framework in a detailed and systematic manner.
We refer in particular to the correct matching and regularization of the 
theory and to the problem of ensuring the UV  finiteness of the bound state 
characteristics and their independence on the the off-shell behaviour of the 
Green functions. 
 
We found it worthwhile to study  these problems by using an appropriate
model. Our results[3], that will appear soon, have proved the possibility of 
developing a consistent quantum field theoretical framework for studying 
bound states with non relativistic lagrangians. 
Our model has a massive scalar bound in an external 
coulombic field and interacting with another massless scalar field. It has the
 advantage of being exactly solvable and it presents all the essential 
features of the physical systems one wants to study, without being affected by
 some inessential complications. 
We have studied this model at different levels of complexity. 

Neglecting, at first stage, the interaction with the massless scalar, 
one computes the relativistic wave functions and  energy levels of the 
heavy scalar in the external field, by looking for the poles of the 
Fourier transform of the two point function. In a non relativistic framework 
one recovers an effective lagrangian starting from the relativistic one and 
requiring that the S matrix elements are the same in both theories. 
Then, one can use Rayleigh-Schr\"odinger perturbation theory to 
evaluate the  energy levels.Our lagrangian, written as an expansion in inverse
 powers of the heavy mass, is built in such a way that the non relativistic 
and relativistic Green functions are the same and this guarantees that the 
correct energy levels are reproduced in the effective theory. The lagrangians 
obtained with different techniques give the same S matrix elements on mass 
shell.
We have shown that they differ only by terms that can be eliminated by using 
field redefinitions and that don't contribute to the bound state energies. 
We have also verified that, even in the effective theories, 
the divergences appearing in the bound state calculations can be cancelled, by summing all the contributions of a given order in the coupling constant and using the same counterterms needed to make the Green functions finite at the same
order.

Finally, we have added an interaction term between the heavy scalar field and a
dynamical massless scalar and computed at first order in the coupling 
constant the energy shift (due to radiative corrections) for the ground state 
of the heavy scalar. 
To perform the non relativistic 
calculation one has to face the problem of the breaking of the naive power 
counting rules. In, fact computing the self energy \footnote{The problem of 
self energy  calculation and of a correct definition of the mass in the 
effective theory will be treated in detail in a separate work [4]} in 
dimensional regularization one sees that the loop contributions fail to 
reproduce the expansion in powers of the momenta on the basis of which the 
effective lagrangian is built. 
So one must impose some additional rules, like the so called ``multipole 
expansion'' that we have adopted, recovering an efficient power counting. 
With this prescription we have verified that the non relativistic 
energy shift at the first order in the coupling constants reproduces exactly 
the relativistic one. 

To summarize,  it is possible to develop a completely consistent and 
well defined formalism, based on the use of non relativistic lagrangians, to 
study the hadronic atoms. These techniques already applied successfully to 
the case of pionium [1-2] could be extended also to the analysis of other 
exotic atoms (like $p\pi^-$, $pK^-$, $d\pi^-$,$dK^-$).

\bigskip

\noindent
[1] P. Labelle and K. Buckley, hep-ph/9804201;
$\,$ H. Kong and F. Ravndall, Phys. Rev. D{\bf{59}}, 014031 (1999); 
hep-ph/9905539; 
$\,$ B. R. Holstein,  nucl-th/9901041;
$\,$ D. Eiras and J. Soto, hep-ph/9905543.\\
\noindent
[2] A. Gall, J. Gasser, V. Lyubovitskij and A. Rusetsky, Phys. Lett. 
B{\bf{462}}, 335 (1999); 
J.~Gasser, V.E.~Lyubovitskij and A.~Rusetsky, 
preprint BUTP-99/20, hep-ph/9910438.\\
\noindent
[3] V. Antonelli, A. Gall, J. Gasser and A. Rusetsky, work in preparation, 
will appear soon. \\
\noindent
[4] V. Antonelli, A. Gall, J. Gasser,  work in progress.

\end{document}